\documentclass[aps,prx,twocolumn,superscriptaddress, longbibliography]{revtex4-2}

\usepackage{graphicx}
\usepackage{dcolumn}
\usepackage{bm}
\usepackage[colorlinks=true,linkcolor=blue,citecolor=blue,urlcolor=blue]{hyperref}

\usepackage{comment, amsmath, mathrsfs, amssymb, braket, xcolor}

\begin{document}

\preprint{APS/123-QED}
\title{Signatures of coherent initial ensembles on all work moments}

\author{Pranay Nayak}
\email{pranay.nayak@fysik.su.se}
\affiliation{Department of Physics, Stockholm University, 106 91, Stockholm, Sweden \\
 }
\author{Sreenath K. Manikandan}%
 \email{skm@tifrh.res.in}
\affiliation{Tata Institute of Fundamental Research Hyderabad, 36/P, Gopanpally Village, Serilingampally Mandal, Hyderabad, Telangana 500046, India}

\author{Tan Van Vu}
\email{tan.vu@yukawa.kyoto-u.ac.jp}
\affiliation{
 Center for Gravitational Physics and Quantum Information, Yukawa Institute for Theoretical Physics, Kyoto University, Kitashirakawa Oiwakecho, Sakyo-ku, Kyoto 606-8502, Japan
}%
\author{Supriya Krishnamurthy$^{1,}$}
\email{supriya@fysik.su.se}
\noaffiliation

\date{\today}

\begin{abstract}
Standard treatments of quantum work using projective energy measurements erase initial coherence and alter the dynamics, thereby failing to capture the thermodynamic effects of coherent superpositions of energy eigenstates in an ensemble of initial states. In this article, we use an operational work definition that is non-intrusive, applying it to the case of a driven dissipative qubit, where the qubit's initial preparation comprises coherent superposition states, while the driving is coherence--less. We derive an evolution equation for the moment generating function for this work, faithfully capturing the thermodynamic signature of coherent superpositions in the initial ensemble. We demonstrate that different initial ensembles that correspond to the same density matrix upon ensemble average, while having the same average work, display different work fluctuations. For monotonic driving, we show that fluctuations are maximum for coherence-less initial ensembles. As an application, we consider quantum bit-erasure in finite time and demonstrate significantly different work statistics for erasing a classical bit of information versus a Haar random initial ensemble. Our results indicate that coherence in the initial ensemble can be utilized as a resource for thermodynamic precision without incurring additional dissipative work costs. We also obtain a generalized fluctuation theorem that establishes a new quantum lower bound on the mean dissipated work. This bound, counterintuitively, is also applicable to a ``classical" initial ensemble with the same initial density matrix and is connected to quantum absolute irreversibility.
\end{abstract}

\maketitle


\section{\label{sec:1}Introduction}

The field of Quantum Thermodynamics attempts to characterize the energetic transformations of driven dissipative quantum processes in terms of the intuitive and familiar notions of heat, work, and entropy production from classical stochastic thermodynamics \cite{campisiColloquiumQuantumFluctuation2011, Vinjanampathy, Potts, Campbell}. A central goal is to design efficient machines that leverage quantum properties, such as coherence \cite{Scully, Scullya, Uzdin, kammerlander_coherence_2016, rodrigues_nonequilibrium_2024, Lostaglioa, Korzekwa, Francica, Lostagliob, Harbola, Abe, Aberg, Goura, Camati, Klatzow} and entanglement \cite{Funob, Hovhannisyanb, Alicki, Brunner, Binder, Perarnau-Llobet, Campaioli, Lobejko}, to create an advantage over their classical counterparts. 

Quantum coherence, specifically the off-diagonal elements of the density matrix in the energy basis (heretofore referred to simply as \textit{coherence}), has been explored as a potential resource at the level of thermodynamic averages \cite{Scullya, Uzdin, kammerlander_coherence_2016, rodrigues_nonequilibrium_2024, Lostaglioa, Korzekwa, Francica, baumgratz_quantifying_2014, Menczel, Watanabe, Francicaa, Lostagliob, santos_role_2019}. However, when analyzing trajectory-wise thermodynamics of small quantum systems weakly coupled to thermal baths, it is known that coherence renders the usual classical definitions of work and heat inapplicable \cite{Talkner, Gallego,  perarnau-llobet_no-go_2017, Baumer2018, elouard_role_2017, Hovhannisyana, Beyera}.
The definition of quantum work, in particular, has been widely investigated.
A meaningful definition of work in the quantum context should still be directly measurable, at least in principle, by an experimentalist manipulating the control parameters. We will call such work definitions operational (see \cite{Baumer2018, dann_unification_2023} for a review of different work definitions).

The most well-studied operational work definition is the Two-Point Measurement (TPM) approach, where projective energy measurements are performed on the system at the start and end of a unitary evolution \cite{Tasaki, Kurchan, Esposito_RMP, Talkner, campisi_colloquium_2011}. This scheme is both well-motivated and intuitive, and incorporating it into the quantum dynamics of the system allows for a full statistical exploration of the role of coherence induced through the Hamiltonian evolution of the system.
It has the desirable feature of reducing to the correct classical notion of work in the absence of coherence, but at the cost of destroying the coherence in both initial and final states as well as intruding on the trajectory evolution \cite{manzano_quantum_2018, garrahan_thermodynamics_2010, Diaz, Solinas, scandi_quantum_2020, miller_joint_2021, van_vu_finite-time_2022}. A very interesting outcome of recent research is the understanding \cite{perarnau-llobet_no-go_2017, Hovhannisyana} that this trade-off is not an isolated scenario, but rather a systemic issue involving all notions of quantum work defined as outcomes of an intrusive Positive Operator-Valued Measure (POVM). No-go theorems \cite{perarnau-llobet_no-go_2017, Hovhannisyana} formalize this fundamental quantum constraint. TPM and other related schemes are hence of limited use, especially when it comes to understanding the role of coherent superposition states in the initial ensemble, which is the primary focus of this paper.

Earlier endeavors to capture the signatures of coherence in the states in the initial ensemble without disturbing it have considered workarounds such as a Bayesian inference of work \cite{micadei_quantum_2020, Micadeib, Parka, Strasberga, rodrigues_nonequilibrium_2024} or quasiprobabilities (work distributions with negative probabilities) \cite{allahverdyan_nonequilibrium_2014, Solinas, miller_time-reversal_2017, Diaz, Francicaa, Gherardini}.
In this paper, we use a quantum work definition that is non-intrusive to the system dynamics \cite{alonso_thermodynamics_2016, elouard_role_2017, naghiloo_heat_2020, Beyera}. It is defined as the expectation value of an operator on the driving apparatus conditioned on some state labels, rather than an outcome of a POVM \cite{Beyera}, and is hence operational while still circumventing the no-go theorems \cite{perarnau-llobet_no-go_2017, Hovhannisyana}. As a result, it captures the effects of coherence both in the coherent initial ensemble and in the driving, while reducing to the classical work for coherence-less trajectories. To our knowledge, no prior study has employed this operational work definition to non-intrusively capture the effect of coherence present in both the initial ensemble and the driving Hamiltonian on work moments. Similar work definitions have, however, been utilized to quantify measurement-induced thermodynamics \cite{elouard_role_2017, elouard_probing_2017, Manikandanb, yanik_thermodynamics_2022}.

As a first step, we consider only the effects of the distribution of states in a coherent initial ensemble on the operational work statistics of a driven dissipative qubit, leaving the case of coherence generated by the driving (also known in the literature as {\it quantum friction}) to a later study. We quantify the effects of this {\em initial coherence} by deriving a closed-form equation for the Moment Generating Function (MGF) of the operational thermodynamic work. All the moments of the work are obtainable from the equation for the MGF, whereby the contributions of the coherence can be easily identified in individual moments. As an application, we show that for optimal average erasure work cost, the presence of coherence in the initial ensemble of states reduces the variance of the work cost distribution. We demonstrate that this variance reduction holds for any monotonic increase in the energy level gap. Unlike the precision bounds of Thermodynamic Uncertainty Relations (TURs) \cite{Barato, Pietzonka, Manikandanc,  Gingrich, Shun, VanVu, VanVue}, this utilization of coherent initial ensembles to achieve precision imposes no additional dissipative cost. Furthermore, we relate the effects of coherence in the initial ensemble to the concept of absolute irreversibility \cite{Murashita, Funoa, Manikandanb}, and quantify this as a correction to Jarzynski equality. Surprisingly, this quantum absolute irreversibility leads to a new bound on the average dissipation even for classical bit erasure. We also obtain a modified Fluctuation-Dissipation Relation (FDR) in the slow driving regime due to coherence in the initial ensemble.

\section{\label{sec:1} Setup}
\subsection{Coherent Initial Ensembles}
To quantify the role of coherent initial ensembles in quantum thermodynamics, we consider a qubit with Hamiltonian $\mathbb{H}_t$. The ground and excited states of the qubit are denoted by $\ket{g}$ and $\ket{e}$, respectively. The initial state of this qubit, $\ket{\psi_0}_i = \mathrm{cos}(\theta/2)\ket{g}+\mathrm{e}^{i\phi}\mathrm{sin}(\theta/2)\ket{e}$, is sampled with probability $\mathrm{Pr\left[ \Theta = \theta, \Phi = \phi \right]}$ having the Cumulative Distribution Function (CDF) 
\[ F_{\psi_0}(\theta, \phi) = \mathrm{Pr}\left[ \Theta \leq \theta, \Phi \leq \phi \right] \]
representing a given ensemble of pure states on the Bloch sphere $\{\ket{\psi_0}_i\}_i$, where  $\theta \in [0,\pi]$ (with random variable $\Theta$) and $\phi \in [0,2\pi)$ (with random variable $\Phi$) are its polar and azimuthal angles, respectively. We refer to a specific state (i.e., a specific combination of $\theta$ and $\phi$ ) by the index $i$. Coherence enters the problem via those states $\ket{\psi_0}_i$ that exist in a superposition of $\ket{g}$ and $\ket{e}$. To decouple the effects of coherent initial ensembles from those of 
quantum friction (drive-induced coherence), we restrict to Hamiltonians that do not generate coherence in the energy basis:
\begin{equation}
\mathbb{H}_t = E_t \ket{e}\!\!\bra{e}.
\end{equation}
Here, $E_t \ge 0$ for $t \in [0,\tau]$ denotes the time-dependent energy gap for a protocol with duration $\tau$. Consequently, any coherence effects observed in the thermodynamics are strictly attributable to the initial CDF $F_{\psi_0}$.

\begin{figure}[t]
\centering
\includegraphics[width=\linewidth]{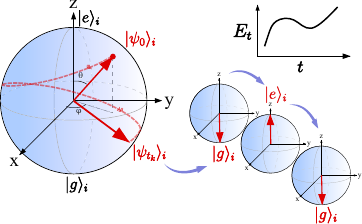}
\caption{\label{fig:1} The schematic depicts a quantum trajectory of a qubit driven with $H_t = E_t \ket{e}\!\!\bra{e}$. The initial state $\ket{\psi_0}_i$ is sampled with a probability $p_i$. It undergoes a non-radiative decay on the surface of the Bloch sphere (red dashed line) until it has its first jump to, say, $\ket{g}$ at time $t_k$ (curved blue arrow). The subsequent quantum jumps between $\ket{g}$ and $\ket{e}$ proceed with Poisson rates dictated by the Kraus operators. The initial state label $i$ identifies the entire trajectory and helps in calculating the correct work statistics, as explained in the text.}
\end{figure}

For later discussions, it is useful to relate the initial ensemble to the density matrix $\rho_0 = \sum_i p_i  \ket{\psi_0}_i\!\!\bra{\psi_0}_i$, which corresponds to the ensemble average over states of the ensemble. Here, $p_i$ denotes the sampling probability of the state $\ket{\psi_0}_i$ in the initial ensemble $\{\ket{\psi_0}_i\}_i$. We would like to compare different coherent initial ensembles. To carry out this comparison meaningfully in our examples, we only choose those \textit{decompositions} that share the same ensemble average $\rho_0$, i.e., different decompositions $\mathscr{D}= \{ p_i, \ket{\psi_0}_i\!\!\bra{\psi_0}_i \}_i$ for a given density matrix $\rho_0$. Further, inspired by Landauer’s symmetric erasure problem, we pick this ensemble average to be $\rho_0 = \mathbb{I}/2$. Note that this ensemble average has no off-diagonal terms. Operationally, the procedure in these examples has the meaning of running the erasure protocol for different initial state preparations of a qubit, which, on average, correspond to a maximally mixed state.

As an example of possible decompositions, the density matrix $\rho_0 = \mathbb{I}/2$ can be decomposed as either $\mathscr{D}_{\rm EG}\equiv \left\{ \{1/2,\ket{e}\!\!\bra{e}\},\{1/2,\ket{g}\!\!\bra{g}\} \right\}$ or $\mathscr{D}_{\rm PM} \equiv \left\{ \{1/2,\ket{+}\!\!\bra{+}\},\{1/2,\ket{-}\!\!\bra{-}\} \right\}$, where $\ket{\pm} = 1/\sqrt{2}(\ket{g}\pm\ket{e})$. We identify thermodynamic signatures of these distinct distributions of coherence in the initial ensemble. However, it is important to note that in order to obtain the work statistics for different decompositions, an experimentalist needs prior information on the state labels $i$ as a necessary and sufficient requirement \cite{Beyera}. This by itself (namely just information about $i$) is not enough to uniquely identify the decomposition $\mathscr{D}$ of the initial ensemble since it is not known which pure state the label $i$ corresponds to. In our example above, both $\mathscr{D}_{\rm EG}$ and $\mathscr{D}_{\rm PM}$ can have states with labels $i = 1,2$, and hence cannot be distinguished by the experimentalist before the work statistics are obtained. This prior labeling requirement also avoids paradoxes that lead to possible super-luminal effects \cite{preskill_lecture_nodate, Peres, Bae, Pati}.

\subsection{Trajectory Heat and Work}
We investigate the quantum effects arising from coherent initial ensembles on the thermodynamics of a driven qubit, weakly coupled to a thermal bath at an inverse temperature $\beta$. The average dynamics of such driven-dissipative quantum systems are well-described by the Markovian Gorini--Kossakowski--Sudarshan--Lindblad (GKSL) master equation \cite{lindblad1976generators, gorini1976completely, albash_quantum_2012, yamaguchi_markovian_2017, dann_time-dependent_2018}, $\dot \rho = \mathscr{L}(\rho)$, where $\rho$ denotes the system's density matrix and $\mathscr{L}$ represents the Lindbladian superoperator. Equivalently, considering the stroboscopic limit over a short time interval $\delta t\ll 1$ ($\hbar = k_B = 1$ throughout the paper), the evolution is described by a dynamical map constructed from time-dependent Kraus operators:
\begin{equation}
\rho(t+ \delta t) = \sum_k \mathbb{M}_k \rho(t) \mathbb{M}_k^\dagger.
\end{equation}
Here, $\{\mathbb{M}_k\}_{k\geq0}$ are Kraus operators satisfying the completeness relation $\sum_k \mathbb{M}_k^\dagger \mathbb{M}_k = \mathbb{I}$. These operators encode both jump and no-jump processes and reflect the choice of the unraveling scheme, which is mathematically defined by the choice of environment basis states. While a Kraus operator representation is non-unique, a physically consistent representation emerges naturally by imposing \textit{detailed balance} for modeling trajectories undergoing thermalization \cite{fagnola2007generators, manzano_nonequilibrium_2015}. The operators encoding stochastic jump events project onto instantaneous eigenstates of $\mathbb{H}_t$. The thermal rates for emission ($\gamma_t (1+\bar{n}_t)$) and absorption ($\gamma_t \bar{n}_t$) are determined by the Bose--Einstein distribution $\bar{n}_t = 1/(\mathrm{e}^{\beta E_t} -1)$ for any bosonic bath. While $\gamma_t$ depends on the specifics of the system-bath interaction, its precise form does not qualitatively alter any of our results. 

The natural Kraus operators for our driven-dissipative qubit are
\begin{equation}
\label{eq:kraus op}
    \mathbb{M}_g = \sqrt{\gamma_t \delta t(1+\bar{n}_t)}\ket{g}\!\!\bra{e},\;  \mathbb{M}_e = \sqrt{\gamma_t \delta t \bar{n}_t}\ket{e}\!\!\bra{g},
\end{equation}
and $\mathbb{M}_0 = \mathbb{I}- i \mathbb{H}_t \delta t - \mathbb{J}/2$ with $\mathbb{J} = \mathbb{M}_g^\dagger \mathbb{M}_g + \mathbb{M}_e^\dagger \mathbb{M}_e$. From the qubit's perspective, these singled out Kraus operators are \textit{emission} $(\mathbb{M}_g)$, \textit{absorption} $(\mathbb{M}_e)$, and \textit{null-event} $(\mathbb{M}_0)$ outcomes. These operators update the state vector stroboscopically at every $\delta t$, yielding physical quantum trajectories---stochastic sequences of pure states conditioned on stroboscopic updates. The \textit{ null-event} accounts for the non-radiative decay of a coherent superposition, describing the non-unitary relaxation toward the ground state (depicted on the Bloch sphere in the schematic in Fig. \ref{fig:1}). As we will see, this is the crucial part of the trajectory that preserves the signatures of the coherent initial states.
An experimental setup mimicking these Kraus operators can be engineered for qubits in thermal or effective thermal environments \cite{Campagne-Ibarcq, Benny, Pekola}, for example, using two-mode squeezed light sources where one mode is used as a probe and the second one used as a reference, as discussed in \cite{Benny}. In Appendix \ref{app: kraus}, we derive these operators from a microscopic model as a proof of concept.

Each quantum trajectory represents a possible realization of the system's evolution and provides a natural framework for defining thermodynamic quantities such as stochastic heat and work \cite{manzano_nonequilibrium_2015, orszagQuantumTrajectories2016, lewalle2020measuring, Vanvua}. The physical trajectory, the time ordered sequence of pure states, $\Gamma = \{\ket{\psi_{n \delta t}}\}_{n\geq 0}$ enables separating the unitary (controlled) and non-unitary (uncontrolled) contributions to the dynamics. This allows us to split the change in the \textit{internal energy} $U(t) \equiv \braket{\psi_t|\mathbb{H}_t|\psi_t}$ of the qubit into two distinct components---work (unitary) and heat (non-unitary). For an interval bounded by two stroboscopic measurements at time $t_n=n\delta t$ and $t_{n+1}=(n+1)\delta t$, the infinitesimal work done is defined as \cite{Maes, alonso_thermodynamics_2016, elouard_role_2017, naghiloo_heat_2020, Beyera}
\begin{equation}
\label{eq:inft work defn}
    \delta W_n = U(t_{n+1}^-) - U(t_n^+),
\end{equation}
and the infinitesimal heat across $t_n$ is 
\begin{equation}
\label{eq:inft heat defn}
    \delta Q_n = U(t_{n}^+) - U(t_n^-).
\end{equation}
Here, $t^\pm_{n}$ denotes the side from which the limit is taken. For our coherence-less Hamiltonian, this reduces to $\delta W_n = \braket{\psi_{t_n^+}| H_{t_{n+1}} - H_{t_n}| \psi_{t_n^+}}$, making the association with unitary energetic changes transparent.
Note that, in the case of a null-event, the non-Hermitian ($\mathbb{J}$) part of the evolution operator causes an energy superposition state to decay to the ground state, hence indirectly contributing to the work statistics through the lowered internal energy in the next time step. 

It has been shown earlier \cite {Beyera} that the work as defined by Eq.~\eqref{eq:inft work defn} is measurable without the knowledge of the system Hamiltonian or the exact initial state. For completeness, we briefly describe the general idea for a closed system (see Ref.~\cite{Beyera} for more details). 
For an arbitrary time-dependent drive $E_t$ on a closed system, the meter which implements this drive is measured by an observable $\Omega_{t_n}$ at every $t_n$ after its interaction with the system. 
It is assumed that the interaction with the system is on such a short time scale that the system {\em does not} entangle with the meter and its effective dynamics is hence unitary. Several runs of this procedure can then be averaged over to obtain an expectation value of the observable $\sum_n\Omega_{t_n}$, which is the work defined as in Eq.~\eqref{eq:inft work defn}. No prior knowledge of the system Hamiltonian is required. However, the experimenter controls the protocol $E_t$ and also has information about the $i$-label of the initial state.
The measurement results $\Omega_{t_n}$ can then be categorized according to this label to estimate the expectation value of work conditioned on the $i$-label of the unknown initial state.
Since this work is not defined as a measurement outcome of a POVM, but rather as an expectation value to which the average $\sum_n \Omega_{t_n}$ measurement outcome converges, it circumvents the no-go theorems \cite{perarnau-llobet_no-go_2017, Hovhannisyana}. This definition can also be extended to the open system scenario \cite{Beyera}.
The definition Eq.~\eqref{eq:inft work defn} does not disturb the coherence between energy eigenstates (it is non-intrusive), unlike other operational work measurement schemes such as the TPM \cite{Talknerb, perarnau-llobetNoGoTheoremCharacterization2017, Hovhannisyana}, making it very suitable for our investigations into the thermodynamic role of coherence in the initial state distribution. We reveal the signatures of this coherent initial distribution in all work moments in the next section.

\begin{figure*}[t]
\centering
\includegraphics[width=\textwidth]{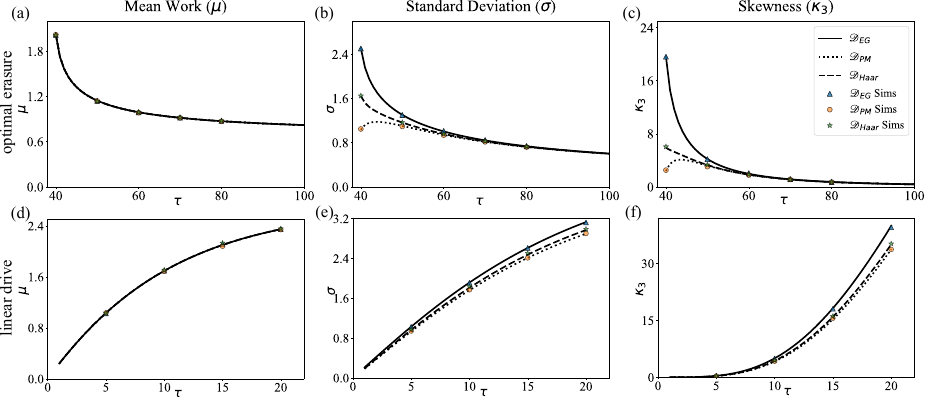}
\caption{\label{fig:mean pm stdev linear} Average work ($\mu =\braket{W}$), standard deviation ($\sigma = \braket{(W-\mu)^2}^{1/2}$), and skewness ($\kappa_3 = \braket{(W-\mu)^3}$) of work cost plotted against total runtime/duration $\tau$. For (a)-(c), an optimal protocol that minimizes mean work cost $\mu$ in a given time interval $[0,\tau]$ is considered, such that the average population in the excited state at $t=0$ is $1/2$, and at $t=\tau$ is $0.01$---this corresponds to optimal erasure with a fixed error of $1\%$. A minimum $\tau$ is required to achieve this erasure fidelity, and different values of $\tau$ correspond to how ``fast'' erasure is carried out. Large $\tau$ corresponds to a slow erasure process. For (d)-(f), a protocol with constant slope $E_t=t/2$ is considered in the interval $[0,\tau]$. The three lines in each figure correspond to three different initial ensembles $\mathscr{D}_{\rm EG}$, $\mathscr{D}_{\rm PM}$, and $\mathscr{D}_{\rm Haar}$. The symbols correspond to the results of Monte Carlo simulations. We choose $\gamma_t = \gamma_0 = 0.1$ and $\beta =1$.}
\end{figure*}

\section{\label{sec: mgf}Work Moment Generating Function}
We consider a protocol $E_t$ which extends over time $0 \leq t \leq \tau$.
The MGF of the work cost is defined as a summation over the ensemble of quantum trajectories $\Gamma$ originating from a given initial decomposition $\mathscr{D}$,
\begin{equation}
\label{eq:path mgf}
    G(u,t) \equiv \sum_{\Gamma } P[\Gamma] \mathrm{e}^{-u W[\Gamma]}.
\end{equation}
Here, we do not explicitly display the dependence of $G(u,t)$ on $\mathscr{D}$, for ease of presentation.
We divide the time $t$ into $N$ stroboscopic steps of size $\delta t \ll 1$ each, intending to take the limit $\delta t \to 0$ later. We will use $t_N$ and $t~(\leq \tau)$ interchangeably. The thermal Kraus operators Eq.~(\ref{eq:kraus op}) map pure states to pure states across these stroboscopic steps. A trajectory of pure states, $\Gamma$, has two parts (as depicted in the schematic in Fig. \ref{fig:1}). That portion of the trajectory until the very first jump occurs,  say at $t_k~(\leq t)$ is referred to as the first part. The second part is simply the rest of the trajectory. In the first part, the system state evolves under the null-event operator $\mathbb{M}_0$. The signature of any initial quantum coherence that the initial state $\ket{\psi_0}$ may have is accounted for through this portion. Here, the state rotates slowly towards the ground state as follows (Appendix \ref{app: mgf}):
\begin{eqnarray}
    \rho_{\rm null}(t_{k}^-) &&=  \ket{\psi_{t_k^-}}\!\!\bra{\psi_{t_k^-}} \nonumber \\ 
    &&= \frac{\mathcal{U}_0(t_k^-,0)  \ket{\psi_0}\!\!\bra{\psi_0} \mathcal{U}_0^\dagger(t_k^-,0) }{\mathrm{Tr}\left[\mathcal{U}_0(t_k^-,0)  \ket{\psi_0}\!\!\bra{\psi_0} \mathcal{U}_0^\dagger(t_k^-,0)\right]}, 
    \label{eq:rho_null}
\end{eqnarray}
where the operator $\mathcal{U}_0(t_k^-,0)$ represents the time ordered sequence of null-events $\mathbb{M}_0 (t_{k-1})  \dots \mathbb{M}_0 (t_{0})$. 
The operational work cost during this time interval $[0,t_k)$ depends on the initial overlap $p_{e,g}^{\psi_0} = \left|\braket{e,g|\psi_0}\right|^2$ and the accumulated thermal decay $K_e(t) \equiv \mathrm{e}^{-\int_0^{t}\gamma_{t'}(\bar n_{t'}+1) dt'}$ and $K_g(t) \equiv \mathrm{e}^{-\int_0^{t}\gamma_{t'}\bar n_{t'} dt'}$, and is given by
\begin{eqnarray}
    W_{\rm null}^{\psi_0}(t_k^-) =&& \int_0^{t_k^-} \mathrm{Tr}\left[   \rho_{\rm null}(t') \dot{\mathbb{H}}(t')\right] dt' \nonumber \\
    =&& \int_0^{t_k^-}   \frac{p_e^{\psi_0} K_e(t')}{p_e^{\psi_0} K_e(t') + p_g^{\psi_0} K_g(t')} \dot{E}_{t'} dt'.
\end{eqnarray}
The coherence-less limit of this truly quantum contribution is $W_{\rm null}^{e}(t_k^-) = E_{t_k^-} - E_0$ (respectively $W_{\rm null}^{g}(t_k^-) = 0$) for an initial state $\ket{e}$ (respectively $\ket{g}$). 
The probability of this null-event trajectory, $P_{\rm null}^{\psi_0}(t_k)$, is simply the denominator of Eq.~(\ref{eq:rho_null}) and can be expressed as
\begin{eqnarray}
    && P^{\psi_0}_{\rm null}(t_k) = \mathrm{Tr}\left[ \mathcal{U}_0(t_k,0)  \ket{\psi_0}\!\!\bra{\psi_0} \mathcal{U}_0^\dagger(t_k,0)  \right] \nonumber \\
    &&= \!\!\!\!  \sum_{j, i \in  \{g,e\}} \!\!\!\!  \bra{j} \left(K_e(t_k) p_e^{\psi_0} \ket{e}\!\!\bra{e} + K_g(t_k) p_g^{\psi_0} \ket{g}\!\!\bra{g} \right) \ket{i} \nonumber \\
     &&\equiv \!\!\!\!  \sum_{j, i  \in  \{g,e\}} \!\!\!\!  \braket{j|\mathbb{P}_{\rm null}^{\psi_0}(t_k)|i}.
\end{eqnarray}
Here, we introduce the matrix $\mathbb{P}_{\rm null}^{\psi_0}$ as a convenient notation that we will use later.  

The second part of the trajectory starts with the first jump (which occurs at time $t_k$) and continues till $t ~(\leq \tau)$. Since, $\mathbb{H}_t$ has time-independent eigenvectors, the operators $\mathbb{M}_g$ or $\mathbb{M}_e$ cause the system to jump to $\ket{g}$ or $\ket{e}$ at time $t_k$. The subsequent trajectory dynamics is a coherence-less jump process between the two eigenstates $\ket{g}$ and $\ket{e}$. In the time interval $[t_{k},t]$, the Kraus operators lead to a time-dependent Poissonian transition rate matrix:
\begin{eqnarray}
    \label{eq:rmatrix}
    \mathbb{R}_t =&&  \gamma_t(\bar{n}_t + 1)\left(  \ket{e}\!\!\bra{e} -  \ket{g}\!\!\bra{e} \right) + \gamma_t\bar{n}_t\left(  \ket{g}\!\!\bra{g} -  \ket{e}\!\!\bra{g} \right), \nonumber \\
\end{eqnarray}
where the transition probability to go from state $\ket{j_n}$ to state $\ket{j_{n+1}}$ at the $n$-th stroboscopic time step is
\begin{equation}
\label{eq:trans prob}
        \braket{j_{n+1}|\mathbb{I} - \mathbb{R}_{t_n} \delta t |j_n}, \quad j_n,j_{n+1} \in \{g, e\}.
\end{equation}
It has been previously observed \cite{Marathe, Imparato, Imparatoa, chvostaProbabilityDistributionWork2007b}, and we outline a proof tailored for our purposes in Appendix \ref{app: mgf}, that in the continuum limit of $\delta t \to 0$, the contribution to the MGF of this effectively classical part of the trajectory starting at a state $\ket{j'}$, $j' \in \{g,e\}$ at time $t_k$ is a time ordered integral
\begin{equation}
\label{eq:class mgf}
    \sum_{ j \in \{g,e\}} \braket{j|\mathcal{T} e^{-\int_{t_k}^{t} (u \mathbb{\dot H} + \mathbb{R}) dt'}|j'},
\end{equation}
where $\mathcal{T}$ denotes the time-ordering operator.
The matrix element $ \braket{j| \mathcal{T}\mathrm{exp}[-\int_{t_{k}}^{t} (u \mathbb{\dot H} + \mathbb{R}) dt']|j'}$ corresponds to a summation over all trajectories that start in state $\ket{j'} \in \{g,e\}$ at time $t_{k}$ and end in state $\ket{j} \in \{g,e\}$ at time $t~(=t_N)$, undergoing one or more jumps in this time interval.

The contribution of all these trajectories to the total MGF can thus be written down as a combination of the contributions of their two parts as
\begin{eqnarray}
     &&\sum_{k \leq N} \sum_{\substack{j, j' \\ \in  \{g,e\}}} \!\!\!\! \braket{j|\mathcal{T}\mathrm{e}^{-\int_{t_k}^{t} (u\dot{\mathbb{H}}_{t'} +\mathrm{\mathbb{R}_{t'}}) dt'} | j'} e^{-u W_{\rm null}^{\psi_0}(t_k^-)} \nonumber \\
     && \quad \quad  \times \mathrm{Tr}\bigg[ \mathbb{M}_{j'}^\dagger (t_k) \mathbb{M}_{j'} (t_k) \mathcal{U}_0(t_k^-,0) \ket{\psi_0}\!\!\bra{\psi_0} \mathcal{U}_0(t_k^-,0)^\dagger \bigg]  \nonumber \\
     &&\to \int_0^{t} \!\! dt_k \!\! \sum_{\substack{j, j', i \\ \in  \{g,e\}}}  \braket{j|\mathcal{T}\mathrm{e}^{-\int_{t_k}^{t}  (u\dot{\mathbb{H}}_{t'} +\mathrm{\mathbb{R}_{t'}}) dt'} | j'} \nonumber \\
     &&\quad \quad \quad \quad \quad \quad \times \bra{j'} \mathbb{T}(t_k) \mathbb{P}_{\rm null}^{\psi_0}(t_k) e^{-u W_{\rm null}^{\psi_0}(t_k)} \ket{i},
\end{eqnarray}
where we have taken the continuous-time limit $\delta t \to 0$ in the second step and replaced the parameters $t_{k}^\pm$ by $t_{k}$ in this limit.
The transition rates for the first jump at $t_k$ are given by the elements of the rate matrix $\mathbb{T}(t_k) = \bar n_{t_k} \gamma_{t_k} \ket{e}\!\!\bra{g} + (\bar n_{t_k} +1) \gamma_{t_k} \ket{g}\!\!\bra{e}$ (i.e., the thermal Poissonian transition rates dictated by $\mathbb{M}_g$ and $\mathbb{M}_e$).

However, there is one trajectory not taken into account above. This is the one, conditioned on the initial state $\ket{\psi_0}$, that  never undergoes a jump in the time interval $[0,t]$, {\em including} at time $t$. Its path probability (again from Eq.~(\ref{eq:rho_null})) is $ P^{\psi_0}_{\rm null}(t)= \sum_{j, i  \in  \{g,e\}}   \braket{j|\mathbb{P}_{\rm null}^{\psi_0}(t)|i}$ and it contributes a work cost $W_{\rm null}^{\psi_0}(t) $.
Hence, the MGF of the work cost upto time $t~( \leq \tau)$ for coherence-less Hamiltonian driving conditioned upon a sampled initial state $\ket{\psi_0}$ can be calculated as 

\begin{equation}
    G^{\psi_0}(u,t)  = \sum_{j, i \in \{g,e\}} \braket{j|\mathbb{G}^{\psi_0}(u,t)|i},
\end{equation}
where the operator $\mathbb{G}^{\psi_0}(u,t)$ is given by
\begin{align}
    \mathbb{G}^{\psi_0}(u,t) &=  \mathbb{P}_{\rm null}^{\psi_0}(t) e^{-u W_{\rm null}^{\psi_0}(t)} \nonumber \\
         &+\int_0^{t} \!\! dt_k  \mathcal{T}\mathrm{e}^{-\int_{t_k}^{t}  (u\dot{\mathbb{H}}_{t'} +\mathrm{\mathbb{R}_{t'}}) dt'}  \nonumber \\
     &\quad  \quad  \times \mathbb{T}(t_k) \mathbb{P}_{\rm null}^{\psi_0} (t_k) e^{-u W_{\rm null}^{\psi_0}(t_k)} .\label{eq:defn gpsi}
\end{align}
The MGF for a given coherent initial ensemble that has CDF $F_{\psi_0}$ is the sum of the elements of the matrix $\mathbb{G}$,
\begin{equation}\label{eq:defn g}
    \mathbb{G}(u,t)= \int \!\!\mathbb{G}^{\psi_0}(u,t) dF_{\psi_0}.
\end{equation}

Formally solving $\mathbb{G}$ from Eq.~\eqref{eq:defn g} is generally difficult. Instead, we take the time derivative of matrix $\mathbb{G}$ to obtain the following differential equation (Appendix \ref{app: mgf}):
\begin{equation}\label{eq:diff quantum mgf}
    \partial_t \mathbb{G}(u,t)  = \left(-u\mathbb{\dot H}(t) -\mathbb{R}(t)\right)\mathbb{G}(u,t)  + \mathbb{C}_{\rm null}(u,t),
\end{equation}
where the operator $\mathbb{C}_{\rm null}(u,t)$ and the initial condition are given by
\begin{equation}
    \mathbb{C}_{\rm null}(u,t) = \int \!\! e^{-u W_{\rm null}^{\psi_0}(t)}u\left( \dot{\mathbb{H}}_t - \dot{W}_{\rm null}^{\psi_0} \right) \mathbb{P}_{\rm null}^{\psi_0} (t) dF_{\psi_0}
\end{equation}
and
\begin{eqnarray}
    \mathbb{G}(u,0) =\! \int\!\! \mathbb{P}_{\rm null}^{\psi_0}(0) dF_{\psi_0} &&= \!\! \int \!\! \left( p_e^{\psi_0}\ket{e}\!\!\bra{e} +p_g^{\psi_0}\ket{g}\!\!\bra{g} \right) dF_{\psi_0} \nonumber\\
    &&\equiv p_e\ket{e}\!\!\bra{e} +p_g\ket{g}\!\!\bra{g}.
\end{eqnarray}

Physically, $\mathbb{C}_{\rm null}$ tracks the contribution to the total work cost statistics of those states in the initial ensemble that are in a superposition of $\ket{g}$ and $\ket{e}$. This contribution from the initial distribution of coherence cannot be captured by the classical MGF Eq.~\eqref{eq:class mgf}. It is obtained from the three possible types of quantum trajectories that never jumped before $t ~(\leq \tau)$. At $t$, they may jump to $\ket{e}$ or $\ket{g}$, or result again in a null-event. Notably, $\mathbb{C}_{\rm null}$ strictly vanishes for initial ensembles with no states in coherent superpositions of energy eigenstates because (for all $\ket{\psi_0}$ in the initial ensemble), either $\dot{W}_{\rm null}^{\psi_0} = \dot E_t$ with $p_g^{\psi_0} = 0$ in $\mathbb{P}_{\rm null}^{\psi_0}$, or $\dot{W}_{\rm null}^{\psi_0} = 0$ with $p_e^{\psi_0} = 0$, rendering the product $\left( \dot{\mathbb{H}}_t - \dot{W}_{\rm null}^{\psi_0} \right) \mathbb{P}_{\rm null}^{\psi_0} (t)$ zero.

\section{\label{sec: results}Results}
\subsection{\label{subsec: heiracrhy}Hierarchical equations of work moments}
We observe that solving Eq.~(\ref{eq:diff quantum mgf}) in the neighborhood of $u=0$ is sufficient for obtaining the moments. By expanding $\mathbb{G}(u,t)$ in powers of $u$ as
\begin{equation}\label{eq:G.expand}
    \mathbb{G}(u,t) = \mathbb{G}_0(t) -u \mathbb{G}_1(t) + \frac{1}{2} u^2 \mathbb{G}_2(t) + \dots,
\end{equation} 
the $n$-th moment of the work cost up to time $t$ is given by $\braket{W^n(t)} = \sum_{j, j' \in \{g,e\}} \bra{j}\mathbb{G}_{n}(t)\ket{j'}$. Substituting Eq.~\eqref{eq:G.expand} to Eq.~\eqref{eq:diff quantum mgf}, we obtain the following hierarchical equations:
\begin{equation}
\label{eq:hierachy_main}
    \partial_t \mathbb{G}_n(t) = - \mathbb{R}(t)\mathbb{ G}_n(t) + n\dot{\mathbb{H}}(t) \mathbb{G}_{n-1}(t) + \mathbb{C}_n(t),
\end{equation}
where the initial conditions are $\mathbb{G}_{n}(0)=0$ for $n\ge 1$ and $\mathbb{G}_0(0) = \mathbb{G}(u,0) = \begin{bmatrix}
    p_e & 0 \\
    0 & p_g
\end{bmatrix}$. The contribution from the initial distribution of coherence is now carried solely by 
\begin{subequations}
\begin{equation}
\label{eq:An mean}
    \mathbb{C}_n(t) = n  \dot{E}(t)  \scalebox{0.9}{$(\ket{g}\bra{g} -\ket{e}\bra{e})$} \! \int (W^{\psi_0}_{\rm null})^{n-1} \mathfrak{a}^{\psi_0} dF_{\psi_0},
\end{equation}
\begin{equation}
\label{eq:a frak mean}
    \mathfrak{a}^{\psi_0}(t) =  p_g^{\psi_0}p_e^{\psi_0} K_g(t)  K_e(t) / P^{\psi_0}_{\rm null}(t).
\end{equation}
\end{subequations}
The term $\mathfrak{a}^{\psi_0}  \propto p_g^{\psi_0} p_e^{\psi_0}$ in $\mathbb{C}_n$ quantifies the effect of coherent initial ensembles since $\mathfrak{a}^{\psi_0}(t) \geq 0$, with equality satisfied \textit{iff} the decomposition is classical, i.e., $\ket{\psi_0}_i \in \{\ket{g}, \ket{e}\}$ for all $i$ labels. 

Notice that $\mathbb{C}_0 = 0$ and $\braket{W(t)}$ is unaffected by $\mathbb{C}_1$ since the  terms of 
$\mathbb{C}_n$ sum to zero for all $n$. Thus, the average work has no true signature of the initial distribution of coherence. The subsequent moments, however, are non-trivially affected (see Fig.~\ref{fig:mean pm stdev linear}).

We plot the non-trivial effects on higher work moments in Fig.~\ref{fig:mean pm stdev linear}, by using Eq.~\eqref{eq:hierachy_main} to calculate the theoretical mean, variance and skewness for two classes of protocols: (i) a simple linear protocol, and (ii) a protocol that optimizes the mean work cost of finite-time bit erasure in runtime $\tau$ with a fixed erasure fidelity of $1\%$, i.e., $\braket{e|\rho_\tau|e}=0.01$. In both cases, we deliberately choose three different initial ensembles that are all represented by the same initial density matrix $\rho_0 = \mathbb{I}/2$---the classical $\mathscr{D}_{\rm EG}$, the quantum $\mathscr{D}_{\rm PM}$, and the Haar ensemble generated by the random Haar Unitary $U_H$ acting on a reference state, generating a uniform distribution of pure states on the Bloch sphere $\mathscr{D}_{\rm Haar}$. The latter two have varying degrees of coherence: one is a discrete distribution, and the other is a continuous distribution on the Bloch sphere. In the corresponding GKSL master equation with density-matrix representation, none of the three ensembles will show any effect due to the coherence for any driving $E_t$. Thus, observing the non-trivial deviations in higher cumulants (variance, skewness, etc.) of total work cost as seen in Fig.~\ref{fig:mean pm stdev linear} requires using a trajectory-wise approach. The effects of coherent initial ensembles are intimately related to the fact that only coherent superpositions undergo a non-radiative decay in the absence of a quantum jump.

\subsection{\label{subsec: coh and fluc}Coherence as a Resource for Precision}
In the symmetric optimal bit erasure setup presented in Fig.~\ref{fig:mean pm stdev linear}, we begin with a degenerate Hamiltonian at $t=0$. {\em Any} initial ensemble with $\rho_0 = \mathbb{I}/2$ has the identical mean cost of erasure for any protocol $E_t$, as mentioned above. The $l_1$ measure for coherence \cite{baumgratz_quantifying_2014} in the initial density matrix $\mathscr{C}_{l_1}(\rho) = \sum_{i,j ( \neq i)} |\rho_{i,j}| $ is also \textit{zero} here, and the driving we consider does not generate coherence. Therefore, the deviations from coherence-less ``classical" $\mathscr{D}_{\rm EG}$ scenario observed in Fig.~\ref{fig:mean pm stdev linear} in the fluctuations (standard deviation $\sigma$) and skewness ($\kappa_3$) of the work cost of bit erasure, are strictly a result of the coherence distribution in the initial ensemble. Moreover, for {\em any} monotonically increasing energy gap $E_t$, we see that fluctuations are maximal for initial ensembles with no coherence. In Eq.~\eqref{eq:flcs and coherence} below, we show that given an initial average population $p_e=\braket{e|\rho_0|e}$ and a monotonically increasing $E_t$, the presence of coherence in the initial ensemble always reduces the variance of the total work cost compared to a corresponding coherence-less ensemble with the same average population. For any time-dependent drive $E_t$ running over a time $t \in [0,\tau]$, the variance of the total work cost $\sigma_W^2(\tau)$ can be obtained using Eq.~\eqref{eq:hierachy_main} (see Appendix \ref{app: coh and fluc} for the details):
\begin{align}
    \sigma_W^2(\tau)&= \sigma^2_{W, \rm cl}(\tau) \nonumber \\
    & \hspace{0.25cm} - 2\int_0^\tau dt\int_0^{t} dt' \dot{E}_t \dot{E}_{t'} \bar{\mathfrak{a}}(t') \mathrm{e}^{-\int_{t'}^t \gamma_{t''} (2\bar{n}_{t''} +1) dt''},\label{eq:flcs and coherence}
\end{align}
where
\begin{align}
    \sigma^2_{W, \rm cl}(\tau) &= 2\int_0^\tau dt  \dot{E}_t \mathrm{e}^{-\int_{0}^t \gamma_{t'} (2\bar{n}_{t'} +1) dt'} \times \nonumber \\
     & \hspace{-0.5cm}\left\{ \braket{W(t)} + \int_0^t dt' \gamma_{t'} \bar{n}_{t'} \braket{W(t')}   \right\}  - \braket{W(\tau)}^2.
\end{align}
Here, $\sigma^2_{W, \rm cl}(\tau)$ is the variance of total work cost for an initial ``classical" ensemble with the same $p_e=\braket{e|\rho_0|e}$, and
$ \mathfrak{\bar{a}} \equiv \int \mathfrak{a}^{\psi_0} dF_{\psi_0} \geq 0$ with equality satisfied \textit{iff} the initial ensemble is coherence-less. 

It is worth noting that the above result differs from earlier results, which have shown that for an invasive TPM work definition with coherence only generated during driving, work fluctuations can increase \cite{scandi_quantum_2020, miller_work_2019}. Our results demonstrate that coherence in the initial ensemble could instead lead to reduced fluctuations and hence are not limited to only being a resource for the average, as considered elsewhere \cite{baumgratz_quantifying_2014, kammerlander_coherence_2016, Francicaa, Lostagliob, santos_role_2019, rodrigues_nonequilibrium_2024, Lostaglioa, Korzekwa, Francica}, but also a resource for the precision of the work cost. An interesting additional point is that 
reducing the fluctuations in our case does not come at an additional average work cost, as opposed to the TUR \cite{Barato, Pietzonka, Manikandanc,  Gingrich, Shun, VanVu, VanVue}.   

\begin{figure}[t]
\centering
\includegraphics[width=\linewidth]{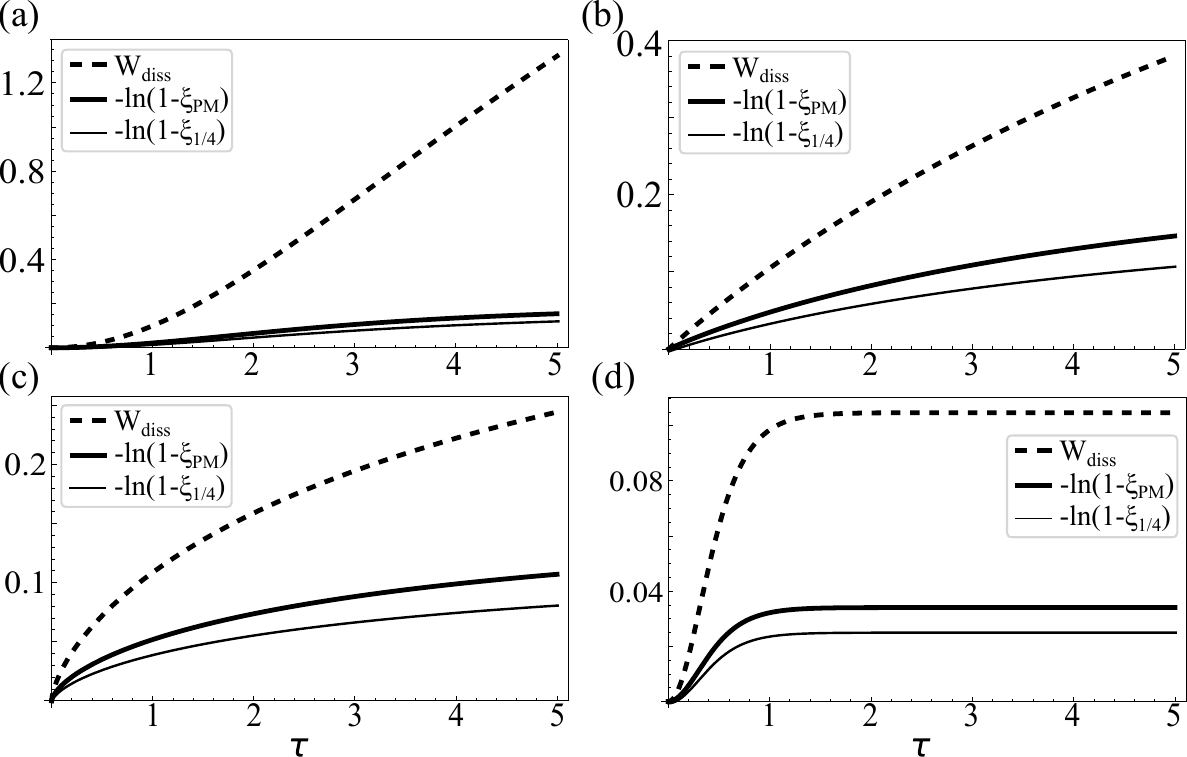}
\caption{\label{fig:wdiss bound}The mean dissipated work $W_{\rm diss}$ (independent of $\mathscr{D}$) is plotted as a function of protocol runtime/duration $\tau$ with an initial ensemble with $p_e=p_g=1/2$ (dashed line) for four different protocols: (a) $E_t = t$, (b) $E_t = t^{1/2}$, (c) $E_t = t^{1/3}$, and (d) $E_t=\mathrm{tanh}(2t)$. The quantum lower bound of $W_{\rm diss}$, $-\ln(1-\xi)/\beta$, is presented for two different coherent decompositions with $\rho_0 = \mathbb{I}/2$. For any given protocol, the mean dissipated work for all the decompositions (including $\mathscr{D}_{\rm EG}$) is the same. $\mathscr{D}_{\rm PM}$ produces the thick line in the figure. The thin line is produced by the equiprobable distribution of $\ket{g}/2 \pm \sqrt{3}\ket{e}/2$. We choose $\gamma_t = \gamma_0 = 0.1$ and $\beta=1$.}
\end{figure}

\subsection{\label{subsec: JE}Modified Jarzynski Equality}
A central result in the field of classical stochastic thermodynamics is the Jarzynski equality (JE) \cite{Jarzynski, Jarzynskia, Crooksa}:
\begin{equation}
\label{eq:Plain JE}
    \braket{\mathrm{e}^{-\beta(W-\Delta F) }} = 1 .
\end{equation}
The average in Eq.~\eqref{eq:Plain JE} is over all system trajectories beginning from an initial equilibrium ensemble (specified by an inverse temperature $\beta$). An amount of work $W$ is performed on the system up to time $\tau$, which typically takes the system out of equilibrium.
At the end of this time duration, the work on the system is stopped, and the system is left to relax to a new equilibrium state at the same $\beta$ value (in this relaxation process, there is only heat dissipated and no extra work performed). The quantity $\Delta F$ is then the difference in free energy between the initial and final equilibrium states. 

Whether Eq.~\eqref{eq:Plain JE} holds or not in the quantum context depends, amongst other things, crucially on the definition of work. For work defined according to the TPM scheme, Eq.~\eqref{eq:Plain JE} holds \cite{Tasaki, Kurchan, Esposito_RMP,Talkner, campisi_colloquium_2011} if the initial ensemble average is 
thermal ($ \rho_0 \propto \mathrm{e}^{-\beta \mathbb{H}_0}$), and the system evolves according to a unital quantum channel \cite{Albash, Rastegin}.
The JE has been further used in the no-go theorem context, to illustrate possible work observables \cite{perarnau-llobet_no-go_2017, Hovhannisyana}.

For schemes that differ from the TPM scheme (in the presence of coherence) and/or involve feedback control, various results \cite{Alhambra, aberg_fully_2018,Francicaa, Beyera, Rubino, Albash, Rastegin, Kwon, Kafri, Solinas,Sagawa,Morikuni, Funo,Venkatesh_2014} suggest that the JE is modified. Another situation where the JE is also modified is linked to the concept of absolute irreversibility in both classical and quantum scenarios \cite{Murashita, Funoa, Murashita_PRA, Manikandanb}. Absolute irreversibility is a notion that arises when some backward trajectories do not have a forward counterpart (or vice versa). This notion is relevant to our case and is apparent when considering trajectories starting from a superposition state. These do not have time-reversed counterparts, since time reversal would require that some backward trajectories jump from an energy eigenstate to a superposition state, which cannot occur under the allowed time evolution. Hence, in analogy with other processes with absolute irreversibility \cite{Murashita, Funoa, Murashita_PRA, Manikandanb}, we expect our qubit system to capture and quantify this effect due to coherent initial states with the parameter $\xi$ defined as
\begin{equation}
\label{eq:JE with xi}
    \braket{\mathrm{e}^{-\beta(W-\Delta F) }} = 1 - \xi.
\end{equation}
Note that $\xi$ must be smaller than or equal to $1$. The average in the LHS of  Eq.~\eqref{eq:JE with xi} is taken over all trajectories $\Gamma$ starting from states in \textit{any} decomposition $\mathscr{D}$ of the density matrix $\rho_0$ (for a degenerate Hamiltonian at $t=0$, $\rho_0 = \mathbb{I}/2$ is the corresponding equilibrium initial density matrix).  Here, $$\Delta F = \int_{E_0}^{E_{\tau}} p_{\rm eq}^e (E) dE $$ is the equilibrium free energy change (not to be confused with the CDF $F_{\psi_0}$) and $p_{\rm eq}^e(E) = 1/(\mathrm{e}^{\beta E} + 1)$ is the isothermal population in the excited state at the energy gap $E$.

We compute $\xi$ by inserting Eq.~\eqref{eq:JE with xi} into the equation for the MGF Eq.~\eqref{eq:diff quantum mgf} with $ u = \beta $. We rewrite the generating equation for the MGF in terms of $\xi$ with a function $\phi(t)$, defined in terms of $E_t$ and $\gamma_t$, and another function $\alpha(t) \geq 0$ that carries the signature of the coherent initial ensemble (see Appendix \ref{app: irreversibility} for exact functional forms):
$$\partial_t^2 \xi +  \phi(t) \partial_t \xi = \alpha(t).$$
Note that $ \alpha(t) > 0$ when coherence is present in the initial ensemble $( \mathfrak{\bar a}(t) >0)$, and $\alpha =0$ for the effectively classical decomposition $\mathscr{D}_{\rm EG}$. Using the initial equilibrium conditions $\xi(0) = \dot \xi(0) =0$, we can write a formal solution for $\xi (t)$, from which it is straightforward to conclude that $\xi(t) \geq 0$. Hence,
\begin{equation}
\label{eq:Jarzynski inequality}
   \braket{\mathrm{e}^{- \beta(W-\Delta F) }} \leq 1.
\end{equation}
The equality is satisfied \textit{iff} the initial decomposition is $\mathscr{D}_{\rm EG}$ (i.e., $\mathfrak{\bar a} = 0 \Rightarrow \xi =0$, see Appendix \ref{app: irreversibility} for more details), wherein all backward trajectories have forward counterparts and the initial state is an equilibrium state \cite{Jarzynski, Searles, Crooksa}. For the operational definition of work used here, and without feedback control \footnote{The modifications to the JE are of a different form with memory and feedback \cite{Sagawa, Morikuni, Funo}. In particular, the LHS of Eq.~\eqref{eq:Jarzynski inequality} can be larger than $1$.}, previous results \cite{Beyera}  have proven the same inequality as Eq.~\eqref{eq:Jarzynski inequality} for a more general setup, using concavity results for operator functions \cite{Beyera}.

\subsection{\label{subsec: bound}A Quantum Bound on Classical Dissipated Work}
Applying Jensen's inequality to Eq.~\eqref{eq:JE with xi}, we obtain the following bound on the dissipated work $W_{\rm diss} = \braket{W} - \Delta F$:
\begin{equation}
    \label{eq:wdiss bound}
    W_{\rm diss} \geq -\beta^{-1}\ln(1-\xi(\tau)).
\end{equation}

For the coherence--less initial decomposition $\mathscr{D}_{\rm EG}$, this lower bound to the mean dissipated work $-\beta^{-1}\ln(1-\xi)$ is \textit{zero}, consistent with the second law of thermodynamics (when beginning from equilibrium initial states). Here, the trajectories and their work values are the same as those of an analogously driven classical Two-Level System (TLS) obeying detailed balance (in this case, the work values also match those obtained from TPM). However, note that {\em all} decompositions of $\rho_0 = \mathbb{I}/2$ yield the same average dissipated work $W_{\rm diss}$ as $\mathscr{D}_{\rm EG}$. But, they can have a non--zero $0<\xi<1$, leading to markedly different lower bounds $-\beta^{-1}\ln(1-\xi)$ to $W_{\rm diss}$, improving upon the quasi--static bound. This discrepancy arises because apart from the particular protocol used, $\xi$ also depends on the specific choice of the decomposition $\mathscr{D}$, while $W_{\rm diss}$ does not. Thus, it appears that the average dissipated work for a classical TLS can be further lower bounded by {\em any} coherent initial decomposition having the same initial ensemble average $\rho_0$. 

In Fig.~\ref{fig:wdiss bound}, we plot $W_{\rm diss}$ against the runtime $\tau$ for four different protocols. For each protocol, we present lower bounds calculated using two different ensembles with coherent decompositions of $\rho_0 = \mathbb{I}/2$. The first is $\mathscr{D}_{\rm PM}$. The second is an ensemble with two states $\ket{\psi_0}_1$ and $\ket{\psi_0}_2$, such that $\left|\braket{e|\psi_0}_1\right|^2 = 1/4$ and $\ket{\psi_0}_2$ is it's polar opposite on the Bloch sphere with $\left|\braket{e|\psi_0}_2\right|^2 = 3/4$.
We see that the tightness of the bound depends on $\xi$, which is non-zero {\em iff} the decomposition $\mathscr{D}$ of $\rho_0$ of the initial ensemble contains states with coherence. Hence, interestingly, this lower bound could, in principle, be optimized for tightness with respect to an initial ensemble with coherence even when used to bound an effectively classical erasure process (as long as both quantum and classical ensembles have the same $\rho_0$). How exactly one could carry out such an optimization is a very interesting question that we leave for future study.

\begin{figure}[t]
\centering

\includegraphics[width=\linewidth]{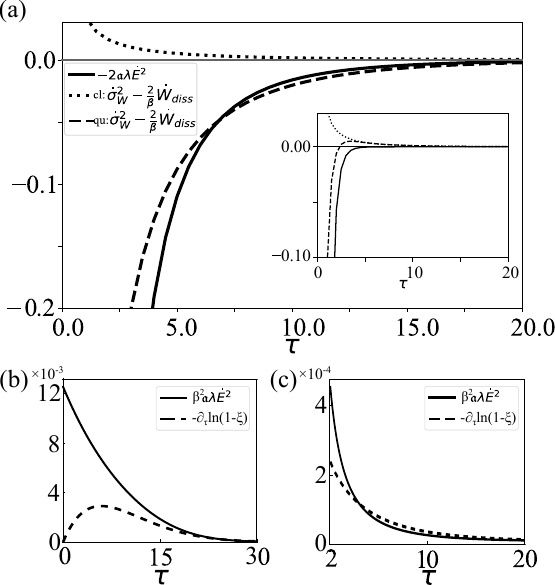}
\caption{In (a), we plot the classical deviation from FDR that goes to zero as runtime $\tau$ increases (cl: $\dot\sigma^2_W - 2\beta^{-1}W_{\rm diss}$ with dotted line). The deviation from FDR due to coherence in the decomposition $\mathscr{D}_{\rm PM}$ (qu: $\dot\sigma^2_W - 2\beta^{-1}W_{\rm diss}$ with dashed line) clearly deviates from the classical behavior. The theoretically predicted deviation from classicality in the slow-driving regime for $\mathscr{D}_{\rm PM}$ is given by the solid line. The drive protocol is $E_t = t/\tau$ with $t \in [0,\tau]$. We chose the so-called Ohmic bath $\gamma_t = 0.1 E_t$ as it retains effects of coherence for large $\tau$. In the inset of (a), we demonstrate with $\gamma_t = \gamma_0 = 0.1$ that the deviation from the classical result and the theoretical prediction all tend to $0$ at long times. For (b) and (c), we consider protocols $E_t=0.1t$ and $E_t = 0.1t^{1/3}$ respectively with runtimes $\tau$. We plot  $-\partial_\tau \ln(1-\xi)$ and compare it with the modification to the FDR $ \beta^2 \mathfrak{\bar{a}} \lambda \dot E^2$ as a function of runtime $\tau$. $\beta=1$ for all the plots.}\label{fig: n2 fdr fluc irr}
\end{figure}

\subsection{Slow Driving: Modified Fluctuation Dissipation Relation}\label{subsec: fdr}
Since the origin of $\xi$ lies in the coherence of the states in the initial ensemble, we expect it to be also related to the reduction of fluctuations as seen in Sec.~\ref{subsec: coh and fluc} and Fig.~\ref{fig:mean pm stdev linear}. In particular, it is of interest to understand the effect of coherent initial ensembles on the variance in the FDR regime. In the absence of coherence (and hence for an effectively classical process), it is known that the distribution of work performed is Gaussian in the slow-driving limit with a variance proportional to the dissipated work $W_{\rm diss}$. This is the celebrated FDR \cite{Onsager, Kubo, Risken, Diezemann}, which is consistent also with the JE. In order to carry out the slow-driving approximation in our case, we rewrite the matrix in Eq.~\eqref{eq:hierachy_main} as follows:
\begin{eqnarray}
    \left[ \mathbb{I} + \lambda \partial_{t} \right] \mathbb{G}_n \ket{\mathrm{S}} = &&\left( p_{\rm eq}^e + \lambda \partial_{t} \right) \braket{ W^n} \ket{e}\nonumber \\
    + &&(1-p_{\rm eq}^e) \braket{ W^n} \ket{g} + \lambda \mathbb{C}_n\ket{\mathrm{S}},
\end{eqnarray}
where $\ket{\mathrm{S}} \equiv\sqrt{2}\ket{+}$ and $\lambda = (1-2p_{\rm eq}^e)/\gamma$. We consider finitely slow changes in $E_t$ in the interval $t \in [0,\tau]$, i.e., $\epsilon \propto \dot{E}_t$ remains small along with $\tau \propto 1/\epsilon$ being large \cite{Mandal}. Parameterizing with respect to $\epsilon$, we consider a series expansion of the $n$-th moment of work as
$$ \braket{W^n(t)} = \Delta F^n + \epsilon f_{n,1} + \epsilon^2 f_{n,2} + \dots.$$
We truncate this expansion up to the second order in $\epsilon$. Within this approximation, Eq.~\eqref{eq:hierachy_main} leads to the following instantaneous relations $\forall n \geq 2$ (Appendix \ref{app: inst FDR}):
\begin{eqnarray}
    \label{eq:wn fdr}
    \partial_t \braket{W^{n}} - n \braket{W^{n-1}}&&  \partial_t \braket{W} \asymp \frac{ n (n-1)}{\beta} \braket{W^{n-2}} \partial_t W_{\rm diss} \nonumber \\
         &&\hspace{0.5cm}- n \braket{g|\mathbb{C}_{n-1}|g}  \lambda \partial_t E. 
\end{eqnarray}
For coherence-less initial decompositions, $\mathbb{C}_{n} = 0$ for all $n$. The resultant family of relations Eq.~\eqref{eq:wn fdr} is equivalent to the usual FDR (Appendix \ref{app: inst fdr reduce to class}):
\begin{equation}
    \label{eq:cums class}
    \dot \sigma^2_W \asymp \frac{2}{\beta} \dot W_{\rm diss} \;\; \text{and} \;\; \dot \kappa_n \asymp 0 \quad \forall n\geq 3,
\end{equation}
where $\kappa_n$ is the $n$-th cumulant of total work cost. However, one finds that coherent initial states, which yield $\mathfrak{\bar a} > 0$, modify the FDR as
 \begin{equation}
    \label{eq:cums n=2 quant}
    \dot \sigma^2_W \asymp \frac{2}{\beta}\dot W_{\rm diss} - 2 \mathfrak{\bar{a}} \lambda \dot E^2.
\end{equation}
Earlier results \cite{miller_work_2019, scandi_quantum_2020} have studied the slow driving modification to the FDR for TPM work schemes, and have shown that the RHS of Eq.~\eqref{eq:cums class} is modified by the addition of a positive quantity \cite{miller_work_2019, scandi_quantum_2020}. In our case, we see instead a reduction in the instantaneous rate of fluctuations for coherent initial ensembles ($\mathfrak{\bar a} >0$) in the FDR regime.
This is consistent also with Eq.~\eqref{eq:flcs and coherence}, which demonstrated that the reduction in the variance (over the classical value) holds for all time,  for all monotonic protocols.

In Fig.~\ref{fig: n2 fdr fluc irr}(a), an initial ensemble with $\rho_0 = \mathbb{I}/2$ is considered undergoing a linear drive of different ``speeds" ($E_t = t/\tau$, where $\tau$ is the runtime). We observe that in the regime of the validity of the classical FDR, the dotted line, represented as cl: $\dot\sigma^2_W - 2\beta^{-1}\dot W_{\rm diss}$ for the classical decomposition $\mathscr{D}_{\rm EG}$ calculated at end time $\tau$, goes to zero, confirming Eq.~\eqref{eq:cums class}. Meanwhile, for the decomposition $\mathscr{D}_{\rm PM}$ with initial coherence, the dashed line, represented as qu: $\dot\sigma^2_W - 2\beta^{-1}\dot W_{\rm diss}$, shows deviations from the classical curve due to an altered variance. This deviation is captured by the thick line representing Eq.~\eqref{eq:cums n=2 quant} obtained from the $\epsilon$ expansion. Modifications to higher cumulants are also obtained as detailed in Appendix \ref{app: coherence inst fdr}, though their $\epsilon$ dependence is missing. For instance, the third and fourth cumulants are
\begin{subequations}
    \begin{equation}
        \partial_t \kappa_3 = 3\mu_{1} a_2 - a_3 \quad \mathrm{and}
    \end{equation}
    \begin{equation}
        \partial_t \kappa_4 +     12 \mu_{1}^2 a_2 - 4\mu_1 a_3  = 6 \mu_{2} a_2 - a_4,
    \end{equation}
\end{subequations}
where $\mu_n = \braket{W^n}$ and $a_n = n\braket{g|\mathbb{C}_{n-1}|g}\lambda \dot E$. The modifications to the cumulants from initial coherence can be inspected with a cumulant expansion of Eq.~\eqref{eq:JE with xi} as 
$$\partial_t \ln(1-\xi) = - \beta\dot W_{\rm diss} + \frac{\beta^2}{2} \dot \sigma_W^2 -\frac{\beta^3}{3!}  \dot\kappa_3 +\dots.$$
If the Gaussianity of the distribution is preserved in the FDR regime, we expect $$\partial_t \ln(1-\xi) \asymp - \beta^2 \mathfrak{\bar{a}} \lambda \dot E^2.$$  Indeed, in Figs.~\ref{fig: n2 fdr fluc irr}(b) and \ref{fig: n2 fdr fluc irr}(c) with two chosen driving protocols ($E_t=0.1t$ and $E_t = 0.1t^{1/3}$), we see that the curves for the rate of change of $\ln\braket{e^{- \beta W_{\rm diss}}}$ (given by $\beta^{-1}\partial_t \ln(1-\xi)$), and the deviation from classical FDR due to initial coherence $- 2 \beta^2\mathfrak{\bar{a}} \lambda \dot E^2$ does converge as the runtime $\tau$ becomes large. A verification with order by order comparision in $\epsilon$ is required, as this could also be due to the decaying probability of having a null--event trajectory with increasing $\tau$. 
In contrast, note that within the TPM scheme and with coherence in driving, it has been verified that the work distribution does not remain Gaussian even in the slow-driving limit \cite{scandi_quantum_2020}.
The connection between $\xi$ and the alteration from classical behavior in the higher moments is left for future studies. 

\section{Summary and Outlook}
In this paper, we have investigated the effects of coherence between energy eigenstates in the initial ensemble on the moments of the thermodynamic work needed to increase the energy gap between the levels of a qubit, a system often studied in the context of (finite-time) information erasure {\cite{browne_guaranteed_2014, elouard_extracting_2017, miller_quantum_2020, van_vu_finite-time_2022}}. Coherence enters only via the states in the initial ensemble, and not due to the driving Hamiltonian. This case has only rarely been studied, since the TPM scheme, which is the scheme most frequently used to define performed work,  erases the effect of any initial coherence. In contrast, our definition of work [Eq.~\eqref{eq:inft work defn}] is chosen to preserve the effects of coherence while also being operational \cite{Beyera}.
This allows us to isolate the specific effects of the coherence distribution in the initial ensemble. We find that for any monotonic driving, coherence in the initial ensemble reduces the variance of the work. Since the initial ensemble can be chosen independently of the driving Hamiltonian, this provides a means to reduce work fluctuations without a trade-off with average work (as opposed to the TURs \cite{Barato, Pietzonka, Gingrich, Manikandanc, Shun, VanVu, VanVue}). 
We demonstrate this reduction of variance in the slow-driving regime with a modified FDR.

In accordance with many earlier works \cite{Alhambra, aberg_fully_2018,Francicaa, Beyera, Rubino, Albash, Rastegin, Kwon, Kafri, Solinas,Sagawa,Morikuni, Funo,Venkatesh_2014}, which have shown that the JE is modified in the quantum context (either in the presence of coherence for an appropriate definition of the thermodynamic work and/or in the context of feedback control), we show that such is the case for the system we have studied.  We quantify the modification to the JE for our setup by calculating the exact deviation induced by the dynamics. The deviation $\xi$ depends on both the driving protocol as well as a measure of the coherence in the initial ensemble. 
From Jensen's inequality, this deviation translates to a non-trivial bound on the mean work. This bound is only non-zero for ensembles with initial coherence, but rather counter-intuitively, it functions as a bound even for classical coherence-less ensembles with the same initial density matrix, since these share the same value of the average work. 
Note that without this procedure, it wouldn't have been possible to obtain this lower bound for an effectively classical process, since the origin of this bound stems from quantum coherence.
A very interesting aspect for future study is to understand how such bounds could be optimized for classical or quantum ensembles
and whether such a procedure can be used to construct bounds on the average work for other systems by ``introducing" absolute irreversibility into their state evolution.

Another very interesting aspect to investigate more closely is the origin of  $\xi$, which modifies the JE in Eq.~\eqref{eq:JE with xi}.
Earlier studies on the generalizations of the JE (for quantum systems) have required it on several different (though not unrelated) grounds such as accounting for the back-action of quantum measurements within a TPM scheme \cite{Kafri},  in the context of quasi-probability work distributions \cite{Francicaa} or operational work definitions \cite{Beyera}, nonunital quantum channels \cite{Rastegin}, modified TPM schemes \cite{Rubino}, generalized measurements \cite{Albash, Venkatesh_2014}, generalized time-reversal operations \cite{Kwon}, constraints on admissible quantum trajectories \cite{Solinas}, or measurements performed directly on a work source \cite{Alhambra, aberg_fully_2018}. 
It would be very interesting to understand the connection of these different approaches to that of absolutely irreversible processes \cite{Murashita, Funoa, Manikandanb, Murashita_PRA}, which also leads to similar generalizations of the JE. In particular, for the case we study here, the deviation term $\xi $ is presumably directly related to the decay of the initial coherence and the probabilities of the resulting system trajectories which do not have a time-reversed counterpart. It would be very illuminating to quantify this connection as well as elucidate further the relations with earlier work. Note that there exist other modifications to the JE, incorporating feedback control in the spirit of Maxwell's demon \cite{Funo, Morikuni, Sagawa} where $\xi$ can be negative, as opposed to our case.

Finally, the case of qudits, where there is more play in the choice of initial ensembles with coherence, could be a fruitful first step to generalizing some of our results on the effects of coherent initial ensembles.  The added role of {\em quantum friction} in this scenario is also an open question. 

\begin{acknowledgments}
We thank Sahar Alipour for pointing out the connection to super-luminal information transfer and comments on an earlier version of the manuscript. We also thank Caio Naves, Jonas Larson, and Henning Kirchberg for illuminating discussions and for posing helpful questions. SKM acknowledges the support from the Department of Atomic Energy, Government of India, under Project Identification No. RTI4007. TVV is supported by JSPS KAKENHI Grant No.~JP23K13032. SK acknowledges the support of the Swedish Research  Council through the grant 2021-05070.
\end{acknowledgments}

\section*{Data Availability}
Mathematica 12 (student edition) codes used can be found on:\\ ``github.com/ThatPranay/Coherent\_Initial\_States.git"

\appendix
\section{\label{app: kraus} Derivation of Thermal Kraus Operators for a Thermal Optical Field}

We consider the following interaction picture Hamiltonian, that conserves the energy between a freely evolving two-level system and an optical radiation field, $H_0(t) = \sum_\nu\hbar \nu \hat{a}_\nu^\dagger \hat{a}_{\nu} + \hbar \omega_t \sigma_z/2$ for $t'\in (t-\delta t/2,t+\delta t/2)$,

\begin{widetext}

    \begin{equation}
    H_{I}(t')=\hbar  \sum_{\nu}G_\nu\left[\sigma_{-}\hat{a}_\nu^\dagger e^{i(\nu-\omega_t)(t'-t+\delta t/2)}+\sigma_{+}\hat{a}_\nu e^{-i(\nu-\omega_t)(t'-t+\delta t/2)}\right],
\end{equation}
where the raising and lowering operators for the two-level system and the field are defined as usual: $\sigma_{+}=|e\rangle\langle g|,~\sigma_{-}=|g\rangle\langle e|$. Irrespective of the exact form of coupling $G_\nu$, we can write the time-evolution operator to leading order as,

\begin{eqnarray}
   e^{-\frac{i}{\hbar}\int_{t-\delta t/2}^{t+\delta t/2} dt' H_{I}(t')} &\approx& \hat{1}-i\left[ \sum_{\nu}G_{\nu}\int_{t-\delta t/2}^{t+\delta t/2}dt' e^{i(\nu-\omega_t)(t'-t+\delta t/2)} \hat{\sigma}_{+}\hat{a}_\nu+\sum_{\nu}G_{\nu}\int_{t-\delta t/2}^{t+\delta t/2}dt' e^{-i(\nu-\omega_t)(t'-t+\delta t/2)}\hat{\sigma}_{-}\hat{a}_\nu^{\dagger}\right]\nonumber\\&-&\frac{1}{2}\bigg(\sum_{\nu,\nu'}G_{\nu}G_{\nu'}\int_{t-\delta t/2}^{t+\delta t/2}dt'\int_{t-\delta t/2}^{t+\delta t/2}dt'' e^{-i(\nu-\omega_t)(t'-t+\delta t/2)}e^{i(\nu'-\omega_t)(t''-t+\delta t/2)}|e\rangle\langle e| \hat{a}_\nu\hat{a}_{\nu'}^{\dagger}\nonumber\\&+&\sum_{\nu,\nu'}G_{\nu'}G_{\nu}\int_{t-\delta t/2}^{t+\delta t/2}dt'\int_{t-\delta t/2}^{t+\delta t/2}dt'' e^{i(\nu-\omega_t)(t'-t+\delta t/2)}e^{-i(\nu'-\omega_t)(t''-t+\delta t/2)}|g\rangle\langle g| \hat{a}_\nu^{\dagger}\hat{a}_{\nu'}\bigg).
\end{eqnarray}
We may now consider that the different modes have randomly distributed phases so that the interference terms average out to zero. We choose therefore  to only keep terms for which $\nu = \nu'$,

\begin{eqnarray}
   e^{-\frac{i}{\hbar}\int_{t-\delta t/2}^{t+\delta t/2} dt' H_{I}(t')} &\approx& \hat{1}-i\left[\sum_\nu G_{\nu}\frac{\exp[i(\nu-\omega_t)\delta t]-1}{i(\nu-\omega_t)} \hat{\sigma}_{+}\hat{a}_\nu+  \sum_\nu G_{\nu}\frac{\exp[-i(\nu-\omega_t)\delta t]-1}{-i(\nu-\omega_t)}\hat{\sigma}_{-}\hat{a}_\nu^{\dagger}\right]\nonumber\\&-&\frac{1}{2}\sum_\nu |G_\nu|^2 \frac{4\sin[\delta t(\nu-\omega_t)/2]^2}{(\nu-\omega_t)^2}\left(|e\rangle\langle e| (\hat{a}_\nu\hat{a}_\nu^{\dagger})+|g\rangle\langle g| \hat{a}_\nu^{\dagger}\hat{a}_{\nu}\right).
\end{eqnarray}
Simplifying the notation,
\begin{equation}
    f_\nu(\delta t)[g(a_{\nu},a_\nu^\dagger)] =  G_{\nu}\frac{\exp[i(\nu-\omega_t)\delta t]-1}{i(\nu-\omega_t)}[g(a_{\nu},a_\nu^\dagger)],
\end{equation}
and
  \begin{equation}
      f_\nu^*(\delta t)f_\nu(\delta t)[g(a_{\nu},a_\nu^\dagger)] = |G_\nu|^2 \frac{4\sin[\delta t(\nu-\omega_t)/2]^2}{(\nu-\omega_t)^2}[g(a_{\nu},a_\nu^\dagger)].
  \end{equation}
With these, we now proceed to identify an operator sum evolution of the two-level atom when the field modes are traced out. The following analysis of the derivation of the operator sum representation closely follows the analysis in Ref.~\cite{Benny}, but is however more general. For the atom in state $|\psi_a\rangle$, and the field in state $|\{\psi_F\}\rangle,$ the state update can be represented in matrix form as,
\begin{eqnarray}
 e^{-\frac{i}{\hbar}\int_{t-\delta t/2}^{t+\delta t/2} dt' H_{I}(t')} |\psi_a\rangle|\{\psi_F\}\rangle\approx\begin{pmatrix}
         1-\sum_\nu f_\nu^* f_\nu\hat{a}_\nu\hat{a}_\nu^{\dagger}/2 &&-i \sum_\nu f_\nu\hat{a}_\nu\\-i\sum_\nu f_\nu^*\hat{a}_\nu^\dagger &&1-\sum_\nu f_\nu^{*}f_\nu\hat{a}_\nu^{\dagger} \hat{a}_\nu/2 
     \end{pmatrix}|\psi_a\rangle|\{\psi_F\}\rangle.
\end{eqnarray}
Similarly, for a generic density matrix $\rho_s\otimes \{\rho_F\}$, for the atom and the field, we may write the state update rule as,
\begin{eqnarray}
\rho_{t+\delta t/2}&\approx&\begin{pmatrix}
         1-\sum_\nu f_\nu^* f_\nu\hat{a}_\nu\hat{a}_\nu^{\dagger}/2 &&-i \sum_\nu f_\nu\hat{a}_\nu\\-i\sum_\nu f_\nu^*\hat{a}_\nu^\dagger &&1-\sum_\nu f_\nu^{*}f_\nu\hat{a}_\nu^{\dagger} \hat{a}_\nu/2 
     \end{pmatrix}\rho_s\otimes \{\rho_F\} \begin{pmatrix}
         1-\sum_\nu f_\nu^* f_\nu\hat{a}_\nu\hat{a}_\nu^{\dagger}/2 &&-i \sum_\nu f_\nu\hat{a}_\nu\\-i\sum_\nu f_\nu^*\hat{a}_\nu^\dagger &&1-\sum_\nu f_\nu^{*}f_\nu\hat{a}_\nu^{\dagger} \hat{a}_\nu/2 
     \end{pmatrix}\nonumber\\
     &=&\hat{\mathcal{M}}_{2}\rho_s\otimes \{\rho_F\}\hat{\mathcal{M}}_{2}^{\dagger}+\hat{\mathcal{M}}_{1}\rho_s\otimes \{\rho_F\}\hat{\mathcal{M}}_{1}^{\dagger}+\hat{\mathcal{M}}_{0}\rho_s\otimes \{\rho_F\}\hat{\mathcal{M}}_{0}^{\dagger}+\text{cross terms}.
\end{eqnarray}
We may now focus on states of the radiation field that are diagonal in the energy basis, such as thermal states. For our thermodynamic considerations, we consider measurements that probe the energy of the field, that is, using projectors $|n\rangle\langle n|$. The unconditional state after the evolution is
\begin{equation}
   \sum_{n} \hat{1}_{2\times 2}|n\rangle\langle n|_{F}\left(\rho_{t+\delta t/2}\right)\hat{1}_{2\times 2}|n\rangle\langle n|_{F} = \hat{\mathcal{M}}_{2}\rho_s\otimes \{\rho_F\}\hat{\mathcal{M}}_{2}^{\dagger}+\hat{\mathcal{M}}_{1}\rho_s\otimes \{\rho_F\}\hat{\mathcal{M}}_{1}^{\dagger}+\hat{\mathcal{M}}_{0}\rho_s\otimes \{\rho_F\}\hat{\mathcal{M}}_{0}^{\dagger}.
\end{equation}
In the above equation, the cross terms have been ignored 
as they project to the diagonal blocks of the combined density matrix with respect to the field. This is equivalent to a rapidly dephasing environment that discards any coherences in the energy basis.  The diagonal terms that remain have the meaning of absorption, emission, and null exchange events in the sense of reference~\cite{Benny}. The corresponding operators are identified as,
\begin{eqnarray}
    \hat{\mathcal{M}}_{1} = \begin{pmatrix}
         0&&0\\-i\sum_\nu f_\nu^{*}\hat{a}_\nu^\dagger  &&0    \end{pmatrix},~~ \hat{\mathcal{M}}_{2} = \begin{pmatrix}
         0 &&-i\sum_\nu f_\nu\hat{a}_\nu\\0&&0
     \end{pmatrix},
\end{eqnarray}

and,

\begin{equation}
 \hat{\mathcal{M}}_{0} = \begin{pmatrix}
         1-\sum_\nu f_\nu^*f_\nu\hat{a}_\nu\hat{a}_\nu^\dagger /2 &&0\\0 &&1-\sum_\nu f_\nu^*f_\nu\hat{a}_\nu^\dagger \hat{a}_\nu/2 
     \end{pmatrix} \approx \begin{pmatrix}
         \sqrt{1-\sum_\nu f_\nu^*f_\nu\hat{a}_\nu\hat{a}_\nu^\dagger} &&0\\0 &&\sqrt{1-\sum_\nu f_\nu^*f_\nu\hat{a}_\nu^\dagger \hat{a}_\nu}
     \end{pmatrix}.
\end{equation}

These operators naturally satisfy the following completeness relation~\cite{Benny}, 

\begin{equation}
    \sum_{i}\text{tr}_{F}[\{\rho_F\}\hat{\mathcal{E}}_i]=\sum_{i}\mathbb{\hat{E}}_i\approx\hat{1}_{2\times 2},
\end{equation}
From $\hat{\mathcal{E}}_i = \hat{\mathcal{M}}_i^\dagger \hat{\mathcal{M}}_i,$ one may identify the  POVM elements for the operator-sum representation as $\hat{\mathbb{E}}_i = \text{tr}_{F}[\rho_F\hat{\mathcal{E}}_i].$ The POVM elements have the following form,
\begin{eqnarray}
    \hat{\mathbb{E}}_{1} = \begin{pmatrix}
        \text{tr}[\{\rho_F\} \sum_{\nu,\nu'}f_\nu^*f_{\nu'} \hat{a}_{\nu'}\hat{a}^\dagger_\nu] &&0\\0&&0
    \end{pmatrix}, \end{eqnarray}
    
    \begin{eqnarray}
        \hat{\mathbb{E}}_{2} = \begin{pmatrix}
          0&&0\\0&&\text{tr}[\{\rho_F\} \sum_{\nu,\nu'}f_{\nu'}^*f_{\nu} \hat{a}^\dagger_{\nu'} \hat{a}_{\nu}]
    \end{pmatrix},
    \end{eqnarray}

and

\begin{eqnarray}
     \hat{\mathbb{E}}_{0} = \begin{pmatrix}
         1-\text{tr}[\{\rho_F\}\sum_\nu f_\nu^*f_\nu\hat{a}_\nu\hat{a}_\nu^\dagger ] &&0\\0 &&1-\text{tr}[\{\rho_F\}\sum_\nu f_\nu^*f_\nu\hat{a}_\nu^\dagger \hat{a}_\nu]
     \end{pmatrix}
\end{eqnarray}
We can now evaluate this for a specific form of the coupling $G_\nu$. Here we consider the standard dipole form as an example,
\begin{equation}
    G_\nu = \left(\frac{\nu}{2\epsilon_0\hbar V}\right)^{1/2}e_{\nu\lambda}.d_{ge}= \left(\frac{\nu}{2\epsilon_0\hbar V}\right)^{1/2}|d_{ge}|\cos\theta,
\end{equation}
where $e_{\nu\lambda}$ is the polarization vector and $d_{ge}$ is the transition dipole moment for the atomic transition.

\begin{eqnarray}
    \sum_\nu f_\nu^*(\delta t)f_\nu(\delta t)[\hat{g}(a_\nu,a_\nu^\dagger)] &=& \sum_\nu |G_\nu|^2 \frac{4\sin[\delta t(\nu-\omega_t)/2]^2}{(\nu-\omega_t)^2}[\hat{g}(a_\nu,a_\nu^\dagger)]\nonumber\\&=&\frac{|d_{ge}|^2}{2\epsilon_0\hbar V}\left(\int_{0}^{\pi}d\theta \sin\theta \cos^2\theta \right)\int_{\omega_t-\delta\nu}^{\omega_t+\delta\nu} \nu D(\nu)\frac{4\sin[\delta t(\nu-\omega_t)/2]^2}{(\nu-\omega_t)^2}[\hat{g}(a_\nu,a_\nu^\dagger)]d\nu \nonumber\\\nonumber\\&=&\frac{|d_{ge}|^2}{2\epsilon_0\hbar V}\left(\frac{2}{3}\right)\int_{\omega_t-\delta\nu}^{\omega_t+\delta\nu} \nu D(\nu)\frac{4\sin[\delta t(\nu-\omega_t)/2]^2}{(\nu-\omega_t)^2}[\hat{g}(a_\nu,a_\nu^\dagger)]d\nu \nonumber\\
    &\approx&\frac{|d_{ge}|^2\omega_t D(\omega_t)[\hat{g}(a_{\omega_t},a_{\omega_t}^\dagger)]}{3\epsilon\hbar V}\int_{\omega_t-\delta\nu}^{\omega_t+\delta\nu}\frac{4\sin[\delta t(\nu-\omega_t)/2]^2}{(\nu-\omega_t)^2}d\nu. 
\end{eqnarray}
Remaining consistent with our assumption of a  large bath with fast decay rates $\delta\nu\delta t\gg 1$, the integral,
\begin{equation}
    \int_{\omega_t-\delta\nu}^{\omega_t+\delta\nu}\frac{4\sin[\delta t(\nu-\omega_t)/2]^2}{(\nu-\omega_t)^2}d\nu\rightarrow\frac{\pi\delta t}{2}.
\end{equation}
Hence we obtain,
\begin{equation}
    \lim_{\delta t\delta\nu\gg 1} f^*(\delta t)f(\delta t) [\cdot]  = \frac{|d_{ge}|^2\omega_t}{3\epsilon_0\hbar V} D(\omega_t)\frac{\pi}{2}\delta t [\cdot]=\frac{|d_{ge}|^2\omega_t}{3\epsilon_0\hbar V} \frac{V\omega_t^2}{\pi^2c^3}\frac{\pi}{2}\delta t  [\cdot]  = \frac{|d_{ge}|^2\omega_t^3}{6\pi\epsilon_0\hbar c^3}\delta t[\cdot]=\gamma_0(\omega_t)\delta t [\cdot],
\end{equation}
where $\gamma_0(\omega_t)$ is the spontaneous emission rate at the energy gap $\omega_t$. Assuming that the optical environment is in a thermal state, we can now identify $\text{tr}\{\rho_F(\omega_t) \hat{a}^{\dagger}_{\omega_t}\hat{a}_{\omega_t}\}=n_{th}(\omega_t),$ as the average thermal population of the optical environment. This yields the following limiting form for the effect operators,

\begin{eqnarray}
    \hat{\mathbb{E}}_{1}  =\begin{pmatrix}
         (n_{th}(\omega_t) +1) \gamma_0(\omega_t)\delta t&&0\\0&&0
    \end{pmatrix}, \end{eqnarray}
    
    \begin{eqnarray}
        \hat{\mathbb{E}}_{2} =\begin{pmatrix}
          0&&0\\0&&n_{th}(\omega_t) \gamma_0(\omega_t)\delta t
    \end{pmatrix},
    \end{eqnarray}

and

\begin{eqnarray}
     \hat{\mathbb{E}}_{0} =\begin{pmatrix}
         1-(n_{th}(\omega_t) +1) \gamma_0(\omega_t)\delta t&&0\\0&&1-n_{th}(\omega_t) \gamma_0(\omega_t)\delta t
    \end{pmatrix}.\nonumber\\
\end{eqnarray}
Using this, we can resolve the effect operators (or POVM elements) to the following operator-sum representation,
\begin{equation}
    \hat{\mathbb{M}}_1 =\hat{\mathbb{M}}_g = \begin{pmatrix}
            0  & 0\\
            \sqrt{(n_{th}(\omega_t)+1)\gamma_0(\omega_t)\delta t }&0
            \end{pmatrix},~~\hat{\mathbb{M}}_2 =\hat{\mathbb{M}}_e= \begin{pmatrix}
            0  & \sqrt{n_{th}(\omega_t) \gamma_0(\omega_t)\delta t  }\\
            0  &0
            \end{pmatrix},\label{mae}
\end{equation}
and,
\begin{eqnarray}
    \hat{\mathbb{M}}_0 = \begin{pmatrix}
         \sqrt{1-(n_{th}(\omega_t) +1) \gamma_0(\omega_t)\delta t  }&&0\\0&&\sqrt{1-n_{th}(\omega_t) \gamma_0(\omega_t)\delta t  }\label{mo}
    \end{pmatrix},
\end{eqnarray}
such that they obey $\hat{\mathbb{E}}_{i}=\hat{\mathbb{M}}_{i}^\dagger\hat{\mathbb{M}}_i$. Using these operators, it can be verified that the unconditional evolution to the leading order obeys the Gorini-Kossakowski-Sudarshan-Lindblad master equation. This example is a demonstration of the emergence of the thermal Kraus operators
used in the text.

\section{\label{app: mgf}Detailed derivation of the MGF of the total work}

The Moment Generating Function (MGF) of total work done till time $t~(\leq \tau)$, given we start from a state $\ket{\psi_0}$ is defined as,
\[G^{\psi_0}(u,t) = \sum_{\Gamma \in \mathscr{F}} P[\Gamma] \mathrm{e}^{-u W[\Gamma]}\].

We will construct the matrix for the Poissonian quantum jumps between the energy eigenstates first---i.e., neither the initial ensemble has coherence, nor the drive. Thus, the trajectory is always in either $\ket{g}$ or $\ket{e}$. The transition probability $\braket{j_N|\mathbb{S}(t)|j0}$ to go from $\ket{j_0}$ to $\ket{j_N}$ in time $t=t_N$ is,
\begin{align}
\label{eq: classical trans prob}
\braket{j_N|\mathbb{S}(t)|j_0} &= \sum_{ \scriptstyle \{ j_1, \dots j_{N-1} \} }^{ \scriptstyle \{g,e\} }  \braket{j_N|\mathbb{I} - \mathbb{R}_{t_{N-1}} \delta t |j_{N-1}}  \nonumber \\
                    &\quad \quad \quad \quad \quad \times \braket{j_{n-1}|\mathbb{I} - \mathbb{R}_{t_{N-2}} \delta t |j_{n-2}} \dots  \braket{j_1|\mathbb{I} - \mathbb{R}_{t_{0}} \delta t |j_0} \nonumber \\
                    &= \bra{j_N}\left[ \mathbb{I} - \mathbb{R}_{t_{N-1}} \delta t \right] \left[ \mathbb{I} - \mathbb{R}_{t_{N-2}} \delta t \right] ... \left[ \mathbb{I} - \mathbb{R}_{t_{1}} \delta t \right] \left[ \mathbb{I} - \mathbb{R}_{t_{0}} \delta t \right]\ket{j_0} \nonumber \\
                    &\to \braket{j_N|\mathcal{T}\mathrm{e}^{-\int_0^t \mathrm{\mathbb{R}(t')} dt'} |j_0} 
\end{align}
where $t_n = n\delta t$, $\mathcal{T}$ is the time ordering operator, and
\[\mathbb{R}_t = \gamma_t(\bar{n}_t + 1)\left(  \ket{e}\bra{e} -  \ket{g}\bra{e} \right) + \gamma_t\bar{n}_t\left(  \ket{g}\bra{g} -  \ket{e}\bra{g} \right),\]
follows from the Kraus operators. The total work for a path $j_0 \to j_1 \to j_2 \dots \to j_{N}$ is then simply,
\[\mathrm{W}(t;~~ j_N, j_{N-1}, \dots j_0) = \sum_{i=1}^N (E_{t_i}^{(j_i)} - E_{t_{i-1}}^{(j_i)}) \]
where, $E_{t_i}^{(j_i)}$ is the energy of the $j_i$-th eigenstate at time $t_i$. \\

For
\begin{equation}
    \zeta_i = \mathrm{e}^{u E_{t_i}^{(e)}} \ket{e}\!\!\bra{e} +  \mathrm{e}^{u E_{t_i}^{(g)}}\ket{g}\!\!\bra{g} ~~\text{and}~~ \mathrm{\mathbb{K}}_i = \mathbb{I} - \mathbb{R}_{t_i} \delta t,
\end{equation}
the MGF for these non-coherent trajectories is,
\begin{align}
    G^{j_0}(u,t) = \braket{\mathrm{e}^{-u W}}_{j_0} &= \sum_{ j_1, \dots j_{N} } \bra{j_N} \left ( \zeta^{-1}_{N} \zeta^{}_{N-1} \mathrm{\mathbb{K}}_{N-1} \right) \ket{j_{N-1}}\!\!\bra{j_{N-1}} \left( \zeta^{-1}_{N-1} \zeta^{}_{N-2} \mathrm{\mathbb{K}}_{N-2}  \right) \dots \nonumber \\
                                  & \quad \quad \times \ket{j_2}\!\!\bra{j_2} \left( \zeta^{-1}_{2} \zeta^{}_{1} \mathrm{\mathbb{K}}_{1} \right) \ket{j_1}\!\!\bra{j_1} \left( \zeta^{-1}_{1}  \zeta_0 \mathrm{\mathbb{K}}_0 \right) \ket{j_0} \\
                                  &=\sum_{ j_N \in \{g,e\}} \braket{j_N| \mathcal{T}\mathrm{e}^{-\int_0^t  (u\dot{\mathbb{H}}_{t'} +\mathrm{\mathbb{R}_{t'}}) \delta t'} |j_0}
\end{align}
This closely follows similar investigations into two level classical system \cite{Marathe, Imparato, Imparatoa, chvostaProbabilityDistributionWork2007b}.

Now we study the probability and the work done by a trajectory that starts from the state $\ket{\psi_0}$, which can be a coherent superposition of $\ket{e}$ and $\ket{g}$. Given our Kraus operators that dictate the state update rules, we obtain the probability of a trajectory that has null-event outcomes till stroboscopic time $t_k = k \delta t$ and then a jump to an energy eigenstate $\ket{j},~~j\in\{e,g\}$, as:
\begin{eqnarray}
     \mathrm{Tr}\left[ \mathbb{M}_j^\dagger (t_k) \mathbb{M}_j (t_k) \mathbb{M}_0 (t_{k-1}) \dots  \mathbb{M}_0 (t_0) \ket{\psi_0}\!\!\bra{\psi_0} \mathbb{M}_0^\dagger (t_0)  \dots \mathbb{M}_0^\dagger (t_{k-1}) \right]
\end{eqnarray}
Right before the jump, the pure state has exponentially decayed towards the ground state as follows:
\begin{eqnarray}
    \rho_{\rm null}(t_{k}^-) =  \ket{\psi_{t_k^-}}\!\!\bra{\psi_{t_k^-}} = \frac{\mathbb{M}_0 (t_{k-1})  \dots \mathbb{M}_0 (t_{0}) \ket{\psi_0}\!\!\bra{\psi_0} \mathbb{M}_0^\dagger (t_0) \dots \mathbb{M}_0^\dagger (t_{k-1})}{\mathrm{Tr}\left[ \mathbb{M}_0 (t_{k-1})  \dots \mathbb{M}_0 (t_{0}) \ket{\psi_0}\!\!\bra{\psi_0} \mathbb{M}_0^\dagger (t_0) \dots \mathbb{M}_0^\dagger (t_{k-1}) \right]}    
\end{eqnarray}
The operational work done during this non-radiative decay for $\mathbb{H}_t = E_t \ket{e}\!\!\bra{e}$ is
\begin{eqnarray}
    W_{\rm null}^{\psi_0}(t_k^-) =&& \int_0^{t_k^-} \mathrm{Tr}\left[   \rho_{\rm null}(t') \dot{\mathbb{H}}(t')\right] dt' \nonumber \\
    =&& \int_0^{t_k^-}   \frac{p_e^{\psi_0} \mathrm{e}^{-\int_0^{t'}\gamma_{t''}(n_{t''}+1) dt''}}{p_e^{\psi_0} \mathrm{e}^{-\int_0^{t'}\gamma_{t''}(n_{t''}+1)dt''} + p_g^{\psi_0} \mathrm{e}^{-\int_0^{t'}\gamma_{t''}n_{t''}dt''}} \dot{E}_{t'} dt' \nonumber \\
    =&& \int_0^{t_k^-}   \frac{p_e^{\psi_0} K_e(t')}{p_e^{\psi_0} K_e(t') + p_g^{\psi_0} K_g(t')} \dot{E}_{t'} dt',
\end{eqnarray}

where $p_e^{\psi_0} = \left|\braket{e|\psi_0}\right|^2$ and $p_g^{\psi_0} = \left|\braket{g|\psi_0}\right|^2$ Thus, we can write the total quantum MGF for the work cost incurred till time $t~ (\leq \tau)$ by summing over all possible lengths of the null-event $t_k ~(\leq t)$ before the first quantum jump occurs to an energy eigenstate $\ket{j}$ with $j \in \{g,e\}$. The subsequent stochastic evolution is a non-coherent two-state process between states $\ket{g}$ and $\ket{e}$ as discussed above.

\begin{align}
\label{eq:quantum total mgf}
G^{\psi_0}(u,t) = &\sum_{j_N, j \in \{g,e\} } \mathrm{Tr}\left[ \mathbb{M}_j^\dagger (t_0) \mathbb{M}_j (t_0)  \ket{\psi_0}\!\!\bra{\psi_0}  \right]  \nonumber \\
         &\quad \quad \quad \times \braket{j_N| \mathcal{T}\mathrm{e}^{-\int_0^t  (u\dot{\mathbb{H}}_{t'} +\mathrm{\mathbb{R}_{t'}}) dt'} | j} \nonumber \\
         &+\sum_{j_N, j} \mathrm{Tr}\left[ \mathbb{M}_j^\dagger (t_1) \mathbb{M}_j (t_1) \mathbb{M}_0 (t_0) \ket{\psi_0}\!\!\bra{\psi_0} \mathbb{M}_0^\dagger (t_0) \right] e^{-u W_{\rm null}^{\psi_0}(t_1^-)} \nonumber \\
         &\quad \quad \quad \times \braket{j_N| \mathcal{T}\mathrm{e}^{-\int_{t_1}^t  (u\dot{\mathbb{H}}_{t'} +\mathrm{\mathbb{R}_{t'}}) dt'} | j} \nonumber \\
         &+ \dots \nonumber \\ \nonumber \\
         &+\sum_{j_N, j} \mathrm{Tr}\left[ \mathbb{M}_j^\dagger (t_k) \mathbb{M}_j (t_k) \mathbb{M}_0 (t_{k-1}) \dots  \mathbb{M}_0 (t_0) \ket{\psi_0}\!\!\bra{\psi_0} \mathbb{M}_0^\dagger (t_0)  \dots \mathbb{M}_0^\dagger (t_{k-1}) \right] e^{-u W_{\rm null}^{\psi_0}(t_k^-)} \nonumber \\
         &\quad \quad \quad \times \braket{j_N|\mathcal{T}\mathrm{e}^{-\int_{t_k}^t  (u\dot{\mathbb{H}}_{t'} +\mathrm{\mathbb{R}_{t'}}) dt'} | j} \nonumber \\
         &+ \dots  \nonumber \\ \nonumber \\
         &+\sum_{j_N, j} \mathrm{Tr}\left[ \mathbb{M}_j^\dagger (t_N) \mathbb{M}_j (t_N) \mathbb{M}_0 (t_{N-1}) \dots \mathbb{M}_0 (t_{0}) \ket{\psi_0}\!\!\bra{\psi_0} \mathbb{M}_0^\dagger (t_0) \dots \mathbb{M}_0^\dagger (t_{N-1}) \right] e^{-u W_{\rm null}^{\psi_0}(t_N^-)} \nonumber \\
         &\quad \quad \quad \times \braket{j_N| j} \nonumber \\
         &+ \mathrm{Tr}\left[ \mathbb{M}_0^\dagger (t_N) \mathbb{M}_0 (t_N) \mathbb{M}_0 (t_{N-1}) \dots \mathbb{M}_0 (t_{0}) \ket{\psi_0}\!\!\bra{\psi_0} \mathbb{M}_0^\dagger (t_0) \dots \mathbb{M}_0^\dagger (t_{N-1}) \right] e^{-u W_{\rm null}^{\psi_0}(t_N^-)} 
\end{align}

To write the MGF as a $2\times 2$ matrix we write the last term in Eq.~(\ref{eq:quantum total mgf}), corresponding to the trajectory that did not undergo any quantum jump event, as 
\begin{align}
     &\sum_{j_N, i\in \{g,e\}} \bra{j_N} \bigg(K_e(t_N) p_e^{\psi_0} \ket{e}\!\!\bra{e} + K_g(t_N) p_g^{\psi_0} \ket{g}\!\!\bra{g} \bigg) e^{-u W_{\rm null}^{\psi_0}(t_N)} \ket{i} \nonumber \\
     &=\sum_{j_N, i\in \{g,e\}} \bra{j_N} \footnotesize\begin{bmatrix}
        K_e(t_N) e^{-u W_{\rm null}^{\psi_0}(t_N)}p_e^{\psi_0} & 0 \\
        0 &  K_g(t_N) e^{-u W_{\rm null}^{\psi_0}(t_N)}p_g^{\psi_0}
    \end{bmatrix} \ket{i} \\
    &=\sum_{j_N, i\in \{g,e\}} \braket{j_N|\mathbb{P}_{\rm null}^{\psi_0}(t_N)e^{-u W_{\rm null}^{\psi_0}(t_N)}|i}.
\end{align}

Henceforth all matrices will be in exactly the same basis as the above matrix. The rest of the trajectories that underwent their first jump at some $t_k ~(\leq t=t_N)$ can be summed together in the following form
\begin{eqnarray}
     &&\sum_{\substack{j_N, j \in  \{g,e\} \\ k \leq N} } \!\!\!\! \braket{j_N|\mathcal{T}\mathrm{e}^{-\int_{t_k}^t  (u\dot{\mathbb{H}}_{t'} +\mathrm{\mathbb{R}_{t'}}) dt'} | j}   \mathrm{Tr}\left[ \mathbb{M}_j^\dagger (t_k) \mathbb{M}_j (t_k) \mathbb{M}_0 (t_{k-1}) \dots  \mathbb{M}_0 (t_0) \ket{\psi_0}\!\!\bra{\psi_0} \mathbb{M}_0^\dagger (t_0)  \dots \mathbb{M}_0^\dagger (t_{k-1}) \right] e^{-u W_{\rm null}^{\psi_0}(t_k^-)} \nonumber \\
     =&&  \sum_{\substack{j_N, j, i \in  \{g,e\} \\ k \leq N} } \!\!\!\! \braket{j_N|\mathcal{T}\mathrm{e}^{-\int_{t_k}^t  (u\dot{\mathbb{H}}_{t'} +\mathrm{\mathbb{R}_{t'}}) dt'} | j} \braket{j|\footnotesize \begin{bmatrix}
    0 &  P_{\to e} (t_k) \\
    P_{\to g} (t_k)  & 0
    \end{bmatrix}  \begin{bmatrix}
    K_e(t_k) p_e^{\psi_0} &  0  \\
    0   & K_g(t_k) p_g^{\psi_0}
    \end{bmatrix} e^{-u W_{\rm null}^{\psi_0}(t_k)} |i} \nonumber \\
    =&& \sum_{j_N, j ,i \in  \{g,e\}} \int_0^t dt_k  \braket{j_N|\mathcal{T}\mathrm{e}^{-\int_{t_k}^t  (u\dot{\mathbb{H}}_{t'} +\mathrm{\mathbb{R}_{t'}}) dt'} | j} \braket{j|\footnotesize \mathbb{T}(t_k) \mathbb{P}_{\rm null}^{\psi_0}(t_k) e^{-u W_{\rm null}^{\psi_0}(t_k)} |i} 
\end{eqnarray}
where transition probabilities $P_{\to e} (t_k) = \bar n_{t_k} \gamma_{t_k} \delta t $ and $P_{\to g} (t_k) = (\bar n_{t_k}+1) \gamma_{t_k} \delta t$ follow the thermal Poissonian transition rates given by $\mathbb{M}_g$ and $\mathbb{M}_e$ for a stroboscopic time $\delta t$. $\mathbb{T}(t_k) = \bar n_{t_k} \gamma_{t_k} \ket{e}\!\!\bra{g} + (\bar n_{t_k} +1) \gamma_{t_k} \ket{g}\!\!\bra{e}$ is same as in the main text.

Hence, we obtain the MGF of total work cost for coherence-less Hamiltonian driving conditioned upon a sampled initial state $\ket{\psi_0}$, $G^{\psi_0}(u,t=t_N)$, as the sum of elements of the matrix $\mathbb{G}^{\psi_0}$. We write down the $\mathbb{G}^{\psi_0}$ for $t ~( \leq \tau)$ in matrix form as follows: 
\begin{eqnarray}
    G^{\psi_0}(u,t)  =&& \sum_{j, i \in \{g,e\}} \braket{j|\mathbb{G}^{\psi_0}(u,t)|i} \nonumber \\
        =&& \sum_{j, i \in  \{g,e\}} \int_0^t dt_k  \bra{j}\mathcal{T}\mathrm{e}^{-\int_{t_k}^t  (u\dot{\mathbb{H}}_{t'} +\mathrm{\mathbb{R}_{t'}}) dt'}  \mathbb{T}(t_k) \mathbb{P}_{\rm null}^{\psi_0}(t_k) e^{-u W_{\rm null}^{\psi_0}(t_k)} \ket{i} \nonumber \\
    &&\quad + \sum_{j, i} \bra{j} \mathbb{P}_{\rm null}^{\psi_0}(t) e^{-u W_{\rm null}^{\psi_0}(t)} \ket{i}
\end{eqnarray}

and the total MGF for the work distribution for a given initial ensemble described by the CDF $F_{\psi_0}$ is,
\begin{equation}
    \mathbb{G}(u,t)= \int_{\ket{\psi_0}} \!\!\mathbb{G}^{\psi_0}(u,t) dF_{\psi_0}.
\end{equation}

Thus matrix $\mathbb{G}(u,t)$ represents the MGF for total work done until a time $t \in [0,\tau]$. To get the differential equation that generates this matrix, we take a derivative of $\mathbb{G}^{\psi_0}(u,t)$ wrt $t$:
\begin{align}
    \partial_{t}\mathbb{G}^{\psi_0} &=  \mathcal{T}\mathrm{e}^{-\int_{t}^t  (u\dot{\mathbb{H}}_{t'} +\mathrm{\mathbb{R}_{t'}}) dt'}  \mathbb{T}(t) \mathbb{P}_{\rm null}^{\psi_0}(t) e^{-u W_{\rm null}^{\psi_0}(t)} + -(u\dot{\mathbb{H}}_{t} +\mathrm{\mathbb{R}_{t}})\mathcal{T}\mathrm{e}^{-\int_{t_k}^t  (u\dot{\mathbb{H}}_{t'} +\mathrm{\mathbb{R}_{t'}}) dt'}  \mathbb{T}(t_k) \mathbb{P}_{\rm null}^{\psi_0}(t_k) e^{-u W_{\rm null}^{\psi_0}(t_k)} \nonumber \\
    &\quad \quad + \partial_t \left[ \mathbb{P}_{\rm null}^{\psi_0}(t) e^{-u W_{\rm null}^{\psi_0}(t)} \right] \nonumber \\
    &=  \mathbb{T}(t) \mathbb{P}_{\rm null}^{\psi_0}(t) e^{-u W_{\rm null}^{\psi_0}(t)}  -(u\dot{\mathbb{H}}_{t} +\mathrm{\mathbb{R}_{t}})\mathcal{T}\mathrm{e}^{-\int_{t_k}^t  (u\dot{\mathbb{H}}_{t'} +\mathrm{\mathbb{R}_{t'}}) dt'}  \mathbb{T}(t_k) \mathbb{P}_{\rm null}^{\psi_0}(t_k) e^{-u W_{\rm null}^{\psi_0}(t_k)} \nonumber \\
    &\quad \quad +\bigg( -u \dot{W}_{\rm null}^{\psi_0} + \footnotesize \begin{bmatrix}
        - \gamma_t (\bar{n}_t + 1) & 0 \\
        0 & - \gamma_t \bar{n}_t
    \end{bmatrix} \bigg) \mathbb{P}_{\rm null}^{\psi_0}(t) e^{-u W_{\rm null}^{\psi_0}(t)} \nonumber \\
    &=   -(u\dot{\mathbb{H}}_{t} +\mathrm{\mathbb{R}_{t}})\mathcal{T}\mathrm{e}^{-\int_{t_k}^t  (u\dot{\mathbb{H}}_{t'} +\mathrm{\mathbb{R}_{t'}}) dt'}  \mathbb{T}(t_k) \mathbb{P}_{\rm null}^{\psi_0}(t_k) e^{-u W_{\rm null}^{\psi_0}(t_k)} \pm  (u\dot{\mathbb{H}}_{t} +\mathrm{\mathbb{R}_{t}}) \mathbb{P}_{\rm null}^{\psi_0}(t) e^{-u W_{\rm null}^{\psi_0}(t)} \nonumber \\
    &\quad \quad +\bigg( -u \dot{W}_{\rm null}^{\psi_0} + \footnotesize \begin{bmatrix}
        - \gamma_t (\bar{n}_t + 1) & 0 \\
        0 & - \gamma_t \bar{n}_t
    \end{bmatrix} + \mathbb{T}(t)\bigg) \mathbb{P}_{\rm null}^{\psi_0}(t) e^{-u W_{\rm null}^{\psi_0}(t)} \nonumber \\
    &=   -(u\dot{\mathbb{H}}_{t} +\mathrm{\mathbb{R}_{t}})\mathbb{G}^{\psi_0}(u,t)  +  (u\dot{\mathbb{H}}_{t} +\mathrm{\mathbb{R}_{t}}) \mathbb{P}_{\rm null}^{\psi_0}(t) e^{-u W_{\rm null}^{\psi_0}(t)} \nonumber \\
    &\quad \quad +\bigg( -u \dot{W}_{\rm null}^{\psi_0} + \mathbb{R}(t)\bigg) \mathbb{P}_{\rm null}^{\psi_0}(t) e^{-u W_{\rm null}^{\psi_0}(t)} \nonumber \\
    &=   -(u\dot{\mathbb{H}}_{t} +\mathrm{\mathbb{R}_{t}})\mathbb{G}^{\psi_0}(u,t)  +  (u\dot{\mathbb{H}}_{t} -u \dot{W}_{\rm null}^{\psi_0}) \mathbb{P}_{\rm null}^{\psi_0}(t) e^{-u W_{\rm null}^{\psi_0}(t)} 
\end{align}

Thus, taking a time derivative of matrix $\mathbb{G}$ results in the following equation:
\begin{subequations}
\label{eq:diff quantum mgf_app}
    \begin{equation}
        \partial_t \mathbb{G}(u,t)  = -\left(u\mathbb{\dot H}(t) +\mathbb{R}(t)\right)\mathbb{G}(u,t)  + \mathbb{C}_{\rm null}(u,t) 
    \end{equation}
    \begin{equation}
        \mathbb{C}_{\rm null}(u,t) = \int_{\ket{\psi_0}} e^{-u W_{\rm null}^{\psi_0}(t)}( u \dot{\mathbb{H}}_t - u \dot{W}_{\rm null}^{\psi_0} ) \mathbb{P}_{\rm null}^{\psi_0}(t)  dF_{\psi_0}
    \end{equation}
\end{subequations}

with 
$$\mathbb{G}(u,0) =\int_{\ket{\psi_0}} \begin{bmatrix}
    p_e^{\psi_0} & 0 \\
    0 & p_g^{\psi_0}
\end{bmatrix} dF_{\psi_0} = \begin{bmatrix}
    p_e & 0 \\
    0 & p_g
\end{bmatrix}$$

\section{\label{app: coh and fluc}Derivation of the variance of the work distribution}

Consider an ensemble of pure states with average excited state population $p_e = \braket{e|\rho_\tau|e}$. One can realize $p_e$ through a whole class of discrete and continuous distributions of pure states on the Bloch sphere. In many cases, these distinct ensembles have the same density matrix representation. We sketch the derivation of the expression for the variance of the total work cost and its dependence on the distribution of coherence in the states of the initial ensemble. The rate of change of variance is given by 
\[\partial_t \sigma_W^2 = \partial_t \langle W^2 \rangle - \langle W \rangle \partial_t \langle W \rangle.\]

Since driving $\mathbb{H}_t = E_t \ket{e}\!\!\bra{e}$ generates no coherence, $\braket{W}  = \int\mathrm{Tr}[\dot{\mathbb{H}}_{t'} \rho_{t'}]dt'$ is the same for any intial ensemble with the same $p_e$. Thus, we focus on the first term $\partial_t \langle W^2 \rangle$. We will use the hierarchical equations from the main text:
\begin{equation}
\label{eq:hierachy}
    \partial_t \mathbb{G}_n(t) = - \mathbb{R}(t)\mathbb{ G}_n(t) + n\mathbb{\dot{H}}(t) \mathbb{G}_{n-1}(t) + \mathbb{C}_n(t).
\end{equation}

In the equation for calculating $\mathbb{G}_2(t)$, we can sum over the matrix elements to obtain

\begin{eqnarray}
    \partial_t \langle W^2 \rangle = \sum_{j,j' \in \{g,e\}} \partial_t \braket{j'|\mathbb{G}_2|j} = 2 \dot{E}_t \left(\braket{e|\mathbb{G}_1|e} + \braket{e|\mathbb{G}_1|g}\right).
\end{eqnarray}

Using the equation for $\mathbb{G}_1(t)$ now,

\begin{equation}
    \partial_t\left(\braket{e|\mathbb{G}_1|e} + \braket{e|\mathbb{G}_1|g}\right) = -\gamma_t (2 \bar n_t+1)\left(\braket{e|\mathbb{G}_1|e} + \braket{e|\mathbb{G}_1|g}\right) + \gamma_t \bar n_t \braket{W} + \dot{E}_t p_e(t) - \braket{g|\mathbb{C}_1|g}.
\end{equation} 
Here, $p_e(t)=\braket{e|\rho_t|e}=\braket{e|\mathbb{G}_0|e} + \braket{e|\mathbb{G}_0|g}$ is the average excited state population as a function of time. Except $\braket{g|\mathbb{C}_1|g}$, everything else is again the same for all the possible ensembles. Now we can rewrite the equation for the variance:

\begin{subequations}
\begin{eqnarray}
    \sigma_W^2(\tau) = \sigma^2_{W, \rm cl}(\tau)  - 2\int_0^\tau dt\int_0^{t} dt' \dot{E}_t \dot{E}_{t'} \bar{\mathfrak{a}}_{t'}\mathrm{e}^{-\int_{t'}^t \gamma (2\bar{n}_{t''} +1) dt''} 
\end{eqnarray}
\begin{eqnarray}
    \sigma^2_{W, \rm cl}(\tau) = 2\int_0^\tau dt  \dot{E}_t \mathrm{e}^{-\int_{0}^t \gamma_{t'} (2\bar{n}_{t'} +1) dt'}  \left\{ \braket{W(t)} + \int_0^t dt' \gamma \bar{n}_{t'} \braket{W(t')}   \right\}  - \braket{W(\tau)}^2
\end{eqnarray}
\end{subequations}

Only $\mathfrak{\bar a} \equiv \int_{\ket{\psi_0}} \mathfrak{a}^{\psi_0} dF_{\psi_0}$, with $\mathfrak{a}^{\psi_0} \propto p_g^{\psi_0} p_e^{\psi_0}$, carries the signature of coherence in those states which have finite support in the state distribution in the initial ensemble. Otherwise, for any `classical' ensemble comprising only $\ket{g}$ and $\ket{e}$ states, $\mathfrak{\bar a} = 0$. The above equations prove that $\forall \dot{E}_t$ which is monotonic,

\begin{equation}
     \sigma_W^2 (\mathfrak{\bar a} =0) >  \sigma_W^2 (\mathfrak{\bar a} >0).
\end{equation}

\section{\label{app: irreversibility}Absolute Irreversibility}

Before proceeding to quantify absolute irreversibility, we will prove Jarzynski's Equality (JE) $\braket{\mathrm{e}^{-\beta W}} =\mathrm{e}^{-\beta \Delta F} $ as a direct consequence of Eq.~(\ref{eq:diff quantum mgf}) with $\mathbb{C}_{\rm null}(u,t)=0$;
\begin{equation}
    \partial_t \mathbb{G}(u,t)  = -\left(u\mathbb{\dot H}(t) +\mathbb{R}(t)\right)\mathbb{G}(u,t).
\end{equation}

Summing over the terms of the matrix,

\begin{equation}
    \partial_t G(u,t)  = -u \dot E_t g(u,t),
\end{equation}

where $g \equiv \braket{e|\mathbb{G}|e} +\braket{e|\mathbb{G}|g} $, and $g(u,t)$ follows the equation

\begin{equation}
    \partial_t g= -u \dot E_t g - \gamma_t (\bar n_t +1)g + \gamma_t \bar n_t (G - g)
\end{equation}

The initial conditions are those of thermal equilibrium of a degenerate Hamiltonian $\mathbb{G}(u,0) = \mathbb{I}/2$, and $\dot E_0 = 0$. One can check that for $u=\beta$, the system of equations has the following solution (satisfying the JE) 
\begin{equation}
    G(u=\beta,t) = \braket{\mathrm{e}^{- \beta W_t}}=\frac{1+\mathrm{e}^{-\beta E_t}}{2}= \mathrm{e}^{-\beta \Delta F}, ~~ \mathrm{and }~~ g = \frac{\mathrm{e}^{- \beta E_t}}{2}.
\end{equation}

where $\Delta F$ is the difference in free energies between the initial and final equilibrium states.


With initial coherence, the system of equations gets modified to
\begin{subequations}
\begin{equation}
    \partial_t \mathbb{G}(u,t)  = -\left(u\mathbb{\dot H}(t) +\mathbb{R}(t)\right)\mathbb{G}(u,t)  + \mathbb{C}_{\rm null}(u,t) 
\end{equation}
    \begin{equation}
    \partial_t G(u,t)  = -u \dot E_t g,
\end{equation}
\begin{equation}
    \partial_t g= -u \dot E_t g - \gamma_t (\bar n_t +1)g + \gamma_t \bar n_t (G - g) + g_{\rm null}
\end{equation}
\end{subequations}

where $g_{\rm null}= \braket{e|\mathbb{C}_{\rm null}|e}$. Note that $\mathbb{C}_{\rm null}$ is diagonal and its elements sum to zero. Without loss of generality, for the solution at $u=\beta$, we choose the ansatz $G(\beta,t) = (1-\xi)(1+\mathrm{e}^{-\beta E_t})/2$ and solve for $\xi$.

The system of equations with quantum coherence becomes the following second-order partial differential equation in $\xi$:
\begin{equation}
    \ddot{\xi} = \phi(t) \dot \xi + \alpha(t),
\end{equation}
with
\[\phi(t) =\frac{ \beta \dot E^2(1-\mathrm{e}^{ \beta E_t}) +  \ddot E (1+ \mathrm{e}^{\beta E_t}) - \dot E \gamma_t \bar n_t(1+\mathrm{e}^{\beta E_t})^2}{(1+\mathrm{e}^{\beta E_t})\dot E }, \]
and the quantum signature is incorporated in the initial ensemble-dependent function:
\[ \alpha(t) = \frac{\beta\dot E_t^2 \mathrm{e}^{\beta E_t}}{(1+\mathrm{e}^{\beta E_t})\dot E } \braket{e|\mathbb{C}_{\rm null}(u=\beta, t)|e} = \frac{\beta^2 \dot E_t^2  K_g(t)}{(1+\mathrm{e}^{ -\beta E_t})} \int_{\ket{\psi_0}} p_g^{\psi_0} p_{e,~null}^{\psi_0}(t) \mathrm{e}^{- W_{\rm null}^{\psi_0}(t)}dF_{\psi_0} \geq 0. \]

The initial conditions for solving this are simply $\xi(0) = 0$ as $E_0 = 0$, and $\dot \xi (0 ) = 0$ as we start from thermal equilibrium at $t=0$, imposing $\dot E_0 = 0$.

Clearly, for no initial coherence, $\alpha(t) = 0$, consequently $\xi(t)$ is {\em zero}, and we recover the JE. In general, for any protocol $E_t$ in the interval $[0,\tau]$, a formal solution for $\xi$ is:
\begin{equation}
    \xi (\tau )  = \int_0^\tau \mathrm{e}^{-\int_0^t \phi(t') dt'} \int_0^{t} \mathrm{e}^{\int_0^{t'} \phi(t'') dt''}\alpha (t') dt' \geq 0.
\end{equation}

Since $G(u,t)$ stays positive by definition, $G(\beta,t) = (1-\xi)(1+\mathrm{e}^{-\beta E_t})/2 \leq (1+\mathrm{e}^{- \beta E_t})/2$, or
\[ \braket{\mathrm{e}^{-\beta W} } \leq \braket{\mathrm{e}^{-\beta \Delta F} }, \]
where the average is taken over all quantum trajectories starting from any decomposition $\mathscr{D}$ (since $H_0$ is degenerate) of the density matrix $\rho_0 = \mathbb{I}/2$.

\section{\label{app: inst FDR}Instantaneous higher order Fluctuation Dissipation Relations}

To derive the fluctuation relations, we start by summing over the elements of the matrix equation (\ref{eq:hierachy}), 

\[ \partial_t  \braket{W^{n+1}}= \sum_{j,j'} \partial_t \braket{j'|\mathbb{G}_{n+1}|j} = (n+1)\dot{E}_t(\braket{e|\mathbb{G}_n|e}+\braket{e|\mathbb{G}_n|g}).\]

Using the above relation and $n_k = p_{\rm eq}^e(t_k)/(1-2p_{\rm eq}^e(t_k))$, where $p_{\rm eq}^e = 1/(\mathrm{e}^{\beta E_t} + 1)$ is the isothermal population in the excited state at time $t$, we rewrite equation (\ref{eq:hierachy}) as:

\begin{align}
    \left[ \mathbb{I} + \frac{1-2p_{\rm eq}^e}{\gamma} \partial_t \right] \mathbb{G}_n &= \begin{bmatrix}
        p_{\rm eq}^e & p_{\rm eq}^e \\
        1-p_{\rm eq}^e & 1-p_{\rm eq}^e
    \end{bmatrix} \mathbb{G}_n + n\frac{1-2p_{\rm eq}^e}{\gamma} \dot{H}_t \mathbb{G}_{n-1}  + \frac{1-2p_{\rm eq}^e}{\gamma} \mathbb{C}_n
\end{align}

Multiplying both sides by $\ket{S}=\sqrt{2}\ket{+} = \ket{g} + \ket{e}$, and using the relations between matrix elements and work moments, we get:

\begin{align}
\label{eq:exact heirarchy re-written}
    \left[ \mathbb{I} + \lambda \partial_t \right] \mathbb{G}_n \ket{+} &= \begin{bmatrix}
        p_{\rm eq}^e + \lambda \partial_t  \\
        1-p_{\rm eq}^e 
    \end{bmatrix} \braket{W^n} + \lambda \mathbb{C}_n\ket{+} 
\end{align}

where $\lambda(E) = (1-p_{\rm eq}^e(E))/\gamma$. Taking the inverse of the operator $ \left[ \mathbb{I} + \lambda \partial_t \right]$ 
\[ \left[ \mathbb{I} + \lambda \partial_t \right]^{-1} = \mathbb{I} - \lambda \dot{E}_t \partial_E + \frac{1}{2}(\lambda \dot{E}_t \partial_E)^2  + \dots \]
    
Furthermore, we assume $\braket{W^n (t)}$ is an analytical function of $\dot E_t$ and $E_t$. Thus,

\[ \braket{W^n} = F^n + \epsilon f_{n,1} + \epsilon^2 f_{n,2} + \dots, \]

where $f_{n,i} = f_{n,i}(E)$ and $\epsilon \propto \dot E_t$ is the expansion parameter. We consider the upper component of equation (\ref{eq:exact heirarchy re-written}):

\begin{eqnarray}
    \left[ \mathbb{I} + \dot E_t \lambda \partial_E \right]\left( \frac{\partial_t\braket{W^{n+1}}}{(n+1)\dot E_t} \right) = \left[ p_{\rm eq}^e + \lambda \dot E_t \partial_E \right] (F^n + \epsilon f_{n,1}  + \dots) - \lambda \braket{g|\mathbb{C}_n|g}
\end{eqnarray}

with $\braket{g|\mathbb{C}_n|g} \propto \epsilon$. For finite but slow driving, parameterized by a small $\epsilon$ and large total time $\tau \propto 1/\epsilon$ we can terminate the series at small orders in $\epsilon$ \cite{Mandal}. Doing so for the first order in $\epsilon$, we obtain:

\begin{eqnarray}
    \frac{\partial_t\braket{W^{n+1}}}{(n+1)\dot E_t} &&= p_{\rm eq}^e \left[ F^n + \epsilon f_{n,1}  \right] +  \lambda \dot E_t \partial_E F^n - \lambda \dot E_t (\partial_E  p_{\rm eq}^e) F^n  - \lambda \dot E_t p_{\rm eq}^e \partial_E F^n - \lambda \braket{g|\mathbb{C}_n|g} + \mathcal{O}(\epsilon^2) \nonumber \\
    &&= \frac{\braket{W^n}\partial_t \braket{W}}{\dot E_t}  + \lambda \dot E_t \partial_E F^n  - \lambda \dot E_t p_{\rm eq}^e \partial_E F^n - \lambda \braket{g|\mathbb{C}_n|g} + \mathcal{O}(\epsilon^2) \nonumber \\
    &&= \frac{\braket{W^n}\partial_t \braket{W}}{\dot E_t}  + \frac{n}{\beta \dot E_t} \braket{W^{n-1}} \partial_t W_{\rm diss} - \lambda \braket{g|\mathbb{C}_n|g} + \mathcal{O}(\epsilon^2) \nonumber \\
\end{eqnarray}

where we used $ p^e = p_{\rm eq}^e - \lambda \dot E_t \partial_E p_{\rm eq}^e +  \mathcal{O}(\epsilon^2)$, $\dot{p}_{\rm eq}^e = -\beta \dot{E}_t p_{\rm eq}^e (1-p_{\rm eq}^e )$ and $\dot{W}_{\rm diss} = - \dot{E}_t^2 \lambda \partial_E p_{\rm eq}^e  +  \mathcal{O}(\epsilon^3) $. We multiply $(n+1)\dot{E}_t$ on both sides to obtain:

\begin{eqnarray}
\label {eq:inst_mom_reln}
   \partial_t \braket{W^{n+1}} - (n+1) \braket{W^{n}} \partial_t \braket{W} \asymp  \frac{ n (n+1) \braket{W^{n-1}}  }{\beta} \partial_t W_{\rm diss} - (n+1) \braket{g|\mathbb{C}_{n}|g}  \lambda \partial_t E. 
\end{eqnarray}

This family of relations is valid for very slow but finite driving. For quasistatic driving, the work probability distribution function is given by $\delta(W-F)$. We consider slow enough driving such that the first correction in the order of driving speed to the quasistatic limit is sufficient to reproduce the statistics.

\section{\label{app: inst fdr reduce to class}Instantaneous moment relations reduce to the classical Gaussian limit}

For the slowly driven coherence-less setup along an interval of length $\tau$, from Eq.~\eqref{eq:inst_mom_reln} above, we obtain the following equations:
\begin{eqnarray}
\label{eq:fdr class n}
         \partial_t \braket{W^{n}} - n \braket{W^{n-1}}  \partial_t \braket{W} &&\asymp  \frac{ n (n-1) \braket{W^{n-2}}  }{\beta} \partial_t W_{\rm diss} \quad \forall n>1.   \\
         &&=\frac{ n (n-1) \braket{W^{n-2}}  }{2} \partial_t \sigma_W^2
\end{eqnarray}

Using the equations for $n=2$, we obtain the Fluctuation Dissipation Relation (FDR) $\beta \partial_t \sigma_W^2 =  2 \partial_t W_{\rm diss}$, or:
$$  \sigma_W^2 =  2 W_{\rm diss}/\beta. $$

It is straightforward to show using eq (\ref{eq:fdr class n}) that the result for $n=3$ implies that the rate of change of the third cumulant,
$\partial_t \kappa_3 = 0$, $\forall t \in [0, \tau]$, or
$$\kappa_3 = 0.$$

We will assume that $\kappa_{n-1} = 0$, and prove that this implies $\kappa_n = 0$, thus proving by induction that Eq.~(\ref{eq:fdr class n}) implies FDR and that all work cumulants higher than two go to zero.

Using the identity for cumulants $\kappa_m$, and moments $\mu_m$,
\begin{equation}
    \label{eq:cums moms}
    \mu_m = \kappa_m + \sum_{l=1}^{m-1} \binom{m-1}{l-1} \kappa_l \mu_{m-l},
\end{equation}
we write the following:

\begin{eqnarray}
    &&\partial_t \mu_n - n \mu_{n-1}\partial_t \kappa_1 \nonumber \\
    &&= \partial_t [\kappa_n + \kappa_1 \mu_{n-1} + (n-1)\kappa_2 \mu_{n-2}] - n[\kappa_1 \mu_{n-2}  + (n-2)\kappa_2 \mu_{n-3} ] \partial_t \kappa_1 \nonumber \\
    &&= \partial_t\kappa_n + \kappa_1 \partial_t \mu_{n-1} + \mu_{n-1}\partial_t \kappa_1   + (n-1)\kappa_2 \partial_t \mu_{n-2} +  (n-1) \mu_{n-2}\partial_t \kappa_2 \nonumber \\
    && \quad \quad \quad \quad \quad  - n\kappa_1^2 \mu_{n-3} \partial_t \kappa_1 - n(n-3)\kappa_1 \kappa_2 \mu_{n-4} \partial_t \kappa_1 - n(n-2)\kappa_2 \mu_{n-3} \partial_t \kappa_1 \nonumber \\
    &&= \partial_t\kappa_n + \kappa_1 [(n-1) \mu_{n-2}\partial_t \kappa_1 + \frac{(n-1)(n-2)}{2}\mu_{n-3}\partial_t \kappa_2 ] + [\mu_{n-2} \kappa_1 + (n-2)\mu_{n-3} \kappa_2] \partial_t \kappa_1  \nonumber \\
    &&\quad \quad \quad + (n-1)\kappa_2 [(n-2)\mu_{n-3}\partial_t \kappa_1 + \frac{(n-2)(n-3)}{2}\mu_{n-4}\partial_t \kappa_2] +  (n-1) [\mu_{n-3} \kappa_1 + (n-3)\mu_{n-3} \kappa_2]\partial_t \kappa_2 \nonumber \\
    && \quad \quad \quad \quad \quad  - n\kappa_1^2 \mu_{n-3} \partial_t \kappa_1 - n(n-3)\kappa_1 \kappa_2 \mu_{n-4} \partial_t \kappa_1 - n(n-2)\kappa_2 \mu_{n-3} \partial_t \kappa_1 \nonumber \\
    &&= \partial_t\kappa_n + n \kappa_1  \mu_{n-2}\partial_t \kappa_1 + \frac{n(n-1)}{2}\mu_{n-3}\kappa_1 \partial_t \kappa_2  \nonumber \\
    &&\quad \quad \quad + \frac{n(n-1)(n-3)}{2}\kappa_2 \mu_{n-4}\partial_t \kappa_2  - n\kappa_1^2 \mu_{n-3} \partial_t \kappa_1 - n(n-3)\kappa_1 \kappa_2 \mu_{n-4} \partial_t \kappa_1  \nonumber \\
    &&= \partial_t\kappa_n + n \kappa_1  [\kappa_1 \mu_{n-3}  + (n-3)\kappa_2 \mu_{n-4}]\partial_t \kappa_1 + \frac{n(n-1)}{2}[\kappa_1 \mu_{n-3}  + (n-3)\kappa_2 \mu_{n-4}] \partial_t \kappa_2  \nonumber \\
    &&\quad \quad \quad - n \kappa_1  [\kappa_1 \mu_{n-3}  + (n-3)\kappa_2 \mu_{n-4}]\partial_t \kappa_1  \nonumber \\
    &&=  \partial_t\kappa_n  + \frac{n(n-1)}{2} \mu_{n-2} \partial_t \kappa_2
\end{eqnarray}
where along with the identity (\ref{eq:cums moms}) for $\mu_n$'s, and relations (\ref{eq:fdr class n}) for $\partial_t \mu_n$'s, we have imposed the induction assumption that $\kappa_{m} =0$ for $m \in [3,n-1]$. Finally, we are left with 
\begin{equation}
    \partial_t \kappa_n = 0 \quad \forall n \geq 3.
\end{equation}
This completes the proof.

\section{\label{app: coherence inst fdr}Slow driving approximation for the higher-order cumulants with coherence }

We have seen that the relation between moments (\ref{eq:fdr class n}) is equivalent to the usual Fluctuation Dissipation relation and also implies the Gaussianity of the total work  distribution. In the presence of coherence, Eq \eqref{eq:inst_mom_reln} becomes:

\begin{eqnarray}
    \partial_t \mu_n - n \mu_{n-1} \partial_t \mu_1 = \binom{n}{2} \mu_{n-2} \partial_t \kappa_2 + \binom{n}{2} \mu_{n-2} a_2 - a_n,
\end{eqnarray}

where, 
\begin{eqnarray}
    a_n \equiv n \braket{g|\mathbb{C}_{n-1}|g}  \lambda \partial_t E.
\end{eqnarray}

Utilizing the identity (\ref{eq:cums moms}) $\forall n \geq 3$, 

\begin{eqnarray}
    &&\partial_t \mu_n - n \mu_{n-1} \partial_t \mu_1 \nonumber \\
    &&= \partial_t \kappa_n +  \sum_{l=1}^{n-1} \binom{n-1}{l-1} \kappa_l \partial_t \mu_{n-l} +  \sum_{l=1}^{n-1} \binom{n-1}{l-1}  \mu_{n-l} \partial_t \kappa_l - n \mu_{n-1} \partial_t \mu_1  \nonumber \\
    &&= \partial_t \kappa_n +  \sum_{l=1}^{n-1} \binom{n-1}{l-1} \kappa_l (n-l)\mu_{n-l-1}\partial_t \mu_1 + \sum_{l=1}^{n-1} \binom{n-1}{l-1} \kappa_l \binom{n-l}{2} \mu_{n-l-2}\partial_t \kappa_2 \nonumber \\
    &&\hspace{1cm}+ \sum_{l=1}^{n-1} \binom{n-1}{l-1} \kappa_l \left( \binom{n-l}{2} \mu_{n-l-2} a_2 - a_l \right) +  \sum_{l=1}^{n-1} \binom{n-1}{l-1}  \mu_{n-l} \partial_t \kappa_l - n \mu_{n-1} \partial_t \mu_1  \nonumber \\
    &&= \partial_t \kappa_n +  (n-1)\mu_{n-1}\partial_t \mu_1  +  \binom{n-1}{2}  \mu_{n-2}\partial_t \kappa_2 \nonumber \\
    &&\hspace{1cm}+ \sum_{l=1}^{n-1} \binom{n-1}{l-1} \kappa_l \left( \binom{n-l}{2} \mu_{n-l-2} a_2 - a_l \right) +  \sum_{l=1}^{n-1} \binom{n-1}{l-1}  \mu_{n-l} \partial_t \kappa_l - n \mu_{n-1} \partial_t \mu_1  
\end{eqnarray}

Thus, we end up with the relation, 

\begin{eqnarray}
    \partial_t \kappa_n + \sum_{l=1}^{n-3} \binom{n-1}{l}\mu_l \partial_t \kappa_{n-l} + \sum_{m=3}^{n-1} \binom{n-1}{m} \kappa_{n-m} \left(\binom{m}{2} \mu_{m-2} a_2 - a_m \right) = \binom{n}{2} \mu_{n-2} a_2 - a_n.
\end{eqnarray}

This can be used to explicitly write the cumulant of any order in terms of the $a_n$s. Starting from 
\begin{subequations}
    \begin{equation}
        \partial_t \kappa_3 = 3\mu_{1} a_2 - a_3,
    \end{equation}
    \begin{equation}
        \partial_t \kappa_4 +     12 \mu_{1}^2 a_2 - 4\mu_1 a_3  = 6 \mu_{2} a_2 - a_4,
    \end{equation}
    \begin{equation}
        \dots \nonumber,
    \end{equation}
\end{subequations}
we get a family of equations.

\end{widetext}

\bibliography{refs0_new}            

\end{document}